\documentclass[aps,prl,twocolumn,groupedaddress,showpacs]{revtex4-1}

\usepackage{amssymb}
\usepackage{amsmath}
\usepackage{graphicx}
\usepackage{enumerate}
\usepackage{mathtools}
\usepackage{subcaption}

\begin{document}


\title{Strange mechanics of the neutrino flavor pendulum}


\author{Lucas Johns}
\email[]{ljohns@physics.ucsd.edu}
\author{George M. Fuller}
\affiliation{Department of Physics, University of California, San Diego, La Jolla, California 92093, USA}


\date{\today}

\begin{abstract}
We identify in the flavor transformation of astrophysical neutrinos a new class of phenomena, a common outcome of which is the suppression of flavor conversion.  Appealing to the equivalence between a bipolar neutrino system and a gyroscopic pendulum, we find that these phenomena have rather striking interpretations in the mechanical picture: in one instance, the gyroscopic pendulum initially precesses in one direction, then comes to a halt and begins to precess in the opposite direction---a counterintuitive behavior that we analogize to the motion of a toy known as a rattleback.  We analyze these behaviors in the early universe, wherein a chance connection to sterile neutrino dark matter emerges, and we briefly suggest how they might manifest in compact-object environments.
\end{abstract}

\pacs{14.60.Pq, 26.35.+c, 97.60.Bw, 26.30.-k}

\maketitle


Neutrino flavor transformation in dense astrophysical environments has incited considerable fervor in recent years, owing to its importance in properly appraising the role of the neutrino sector in such disparate arenas as dark matter \cite{shi1999, abazajian2001b, asaka2005, asaka2005b, kishimoto2008, petraki2008, bezrukov2013, venumadhav2016, adhikari2017}, baryogenesis \cite{akhmedov1998, abada2006, nardi2006, blanchet2007, shaposhnikov2008, canetti2013}, weak decoupling and Big Bang nucleosynthesis \cite{dolgov2002, abazajian2002, wong2002, simha2008, mangano2011, mangano2012, pastor2009, gava2010}, and dynamics and nucleosynthesis in supernovae and compact-object mergers \cite{fuller1987, fuller1992, qian1995a, qian1995b, qian1997, horowitz1999, mclaughlin1999, schirato2002, pastor2002, pastor2002b, fetter2003, duan2006b, duan2011, tamborra2012, malkus2012, wu2015}.  Moreover, the flavor structure of a Galactic core-collapse supernova neutrino burst is of interest for terrestrial detectors aimed at understanding not only the physics of the source but also the fundamental properties of neutrinos themselves \cite{dighe2000, abazajian2011, scholberg2017}. Far from providing neat resolution, the prolonged siege on this topic has instead continued to reveal facets of the problem that may solicit new conceptual and computational paradigms altogether \cite{sawyer2005, cherry2012, raffelt2013, hansen2014, vlasenko2014b, serreau2014, abbar2015, mirizzi2015, keister2015, armstrong2016, chakraborty2016, sawyer2016, volpe2016, johns2017, dasgupta2017, tian2017, wu2017, izaguirre2017, cirigliano2017}.

\begin{figure}
\includegraphics[width=.46\textwidth]{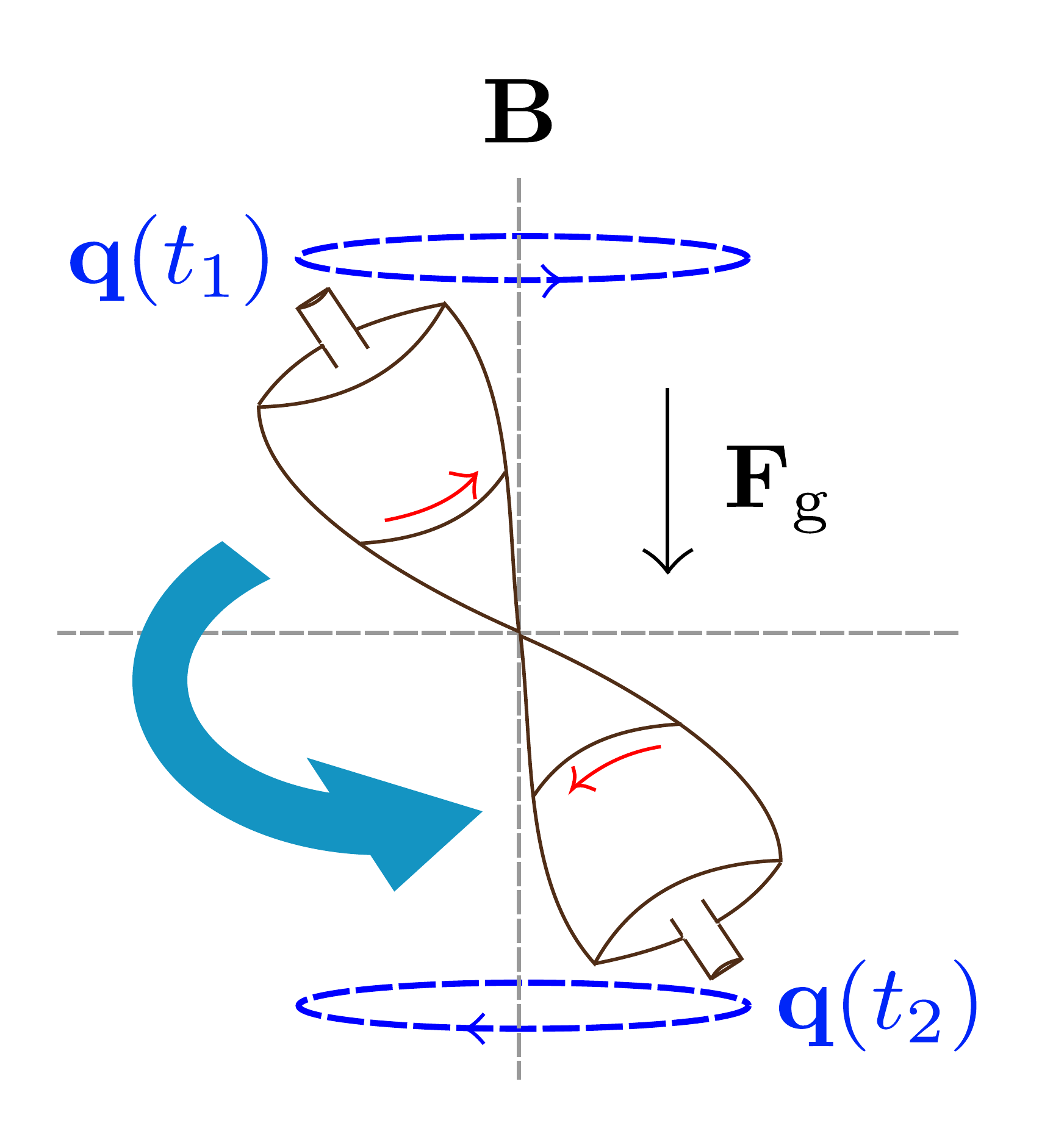}
\caption{One of the strange behaviors exhibited by certain bipolar neutrino systems.  In going from an early time $t_1$ to a late time $t_2$, the flavor pendulum $\mathbf{q}$ falls from its upright position and reverses its spin, hence also the handedness of its precession.}  
\label{top}
\end{figure}

In what follows, we revisit the surprising equivalence that exists between an astrophysical neutrino system, treated in a certain limit, and the mechanical system of a gyroscopic pendulum (\textit{i.e.}, a spinning top whose axis of rotation can swing freely under gravity) \cite{hannestad2006, duan2007b}.  In this mapping between systems, the pendulum can swing, spin, and precess---and each of these motions corresponds to a change in the flavor content of the neutrino population.  We show here that under appropriate circumstances the neutrino system also exhibits some previously unidentified behaviors with rather startling mechanical analogues.  In one case, the spin of the top reverses, leading to a flip in the handedness of precession (Fig.~\ref{top}).  In another, the precession frequency of the top falls precipitously until reaching a critical threshold, after which it hastily speeds up again.  Beyond being counterintuitive, these phenomena may in fact be significant in such environments as the early universe or core-collapse supernovae.

The fact that background particles can dramatically alter neutrino flavor transformation is well known, having played a pivotal part in the solution to the solar neutrino puzzle.  But whereas the Mikheyev--Smirnov--Wolfenstein (MSW) mechanism \cite{wolfenstein1978, mikheyev1985} in the Sun is driven by the forward scattering of neutrinos on electrons, flavor conversion in the inner region of a core-collapse supernova (which has a neutrino number density \textit{$\sim 25$ orders of magnitude} greater than that found in the solar interior) is a nonlinear dynamical problem in which the quantum flavor states of all neutrinos on intersecting trajectories are coupled together by virtue of neutrino--neutrino forward scattering.  This yoking-together of neutrinos gives rise to a host of flavor-transformation phenomena, many radically different from the classic MSW effect, which are grouped together under the epithet \textit{collective oscillations} \cite{duan2010, chakraborty2016}.

In the two-flavor approximation, the flavor content of neutrinos (antineutrinos) of a given energy is customarily written as a polarization vector $\mathbf{P}$ ($\bar{\mathbf{P}}$), where the projection onto the $z$-axis gives the difference in number densities of the two flavors.  One striking example of collective oscillations is the phenomenon of bipolar flavor transformation, in which the evolution of the system is captured by two interacting blocks of polarization vectors, one representing neutrinos of all energies, the other antineutrinos of all energies \cite{kostelecky1995, samuel1996}.  In the absence of a matter background (\textit{e.g.}, charged leptons), the equations of motion in the bipolar regime are
\begin{align}
& \dot{\mathbf{P}} = \left( +\omega \mathbf{B} - \mu \bar{\mathbf{P}} \right) \times \mathbf{P}, \notag \\
& \dot{\bar{\mathbf{P}}} = \left( -\omega \mathbf{B} + \mu \mathbf{P} \right) \times \bar{\mathbf{P}}, \label{ppbar}
\end{align}
where the dot denotes a time derivative and $\mathbf{B} = \pm \left( \sin 2\theta, 0, - \cos 2\theta \right)$, with $\theta$ the vacuum mixing angle.  The choice of plus (minus) corresponds to the normal (inverted) neutrino mass hierarchy.  The oscillation frequency $\omega$ is an average of the oscillation frequencies $\omega_i$ of the individual neutrinos (labelled by index $i$), which in vacuum are $\omega_i = | \delta m^2 | / 4 E_i$, where $E_i$ designates the neutrino energy.  Lastly, $\mu$ is the potential generated by neutrino--neutrino forward scattering.  In the early-universe calculations that follow we use a normalization such that $\mu = \sqrt{2} G_F T^3$, with $G_F$ the Fermi coupling constant and $T$ the temperature.  For the time being, we postpone discussion of the matter background. Throughout this study we neglect non-forward scattering (\textit{i.e.}, collisions); see the Supplemental Material for justification.

To see the equivalence of this system to a gyroscopic pendulum, we introduce the vectors $\mathbf{D} = \mathbf{P} - \bar{\mathbf{P}}$ and $\mathbf{Q} = \mathbf{P} + \bar{\mathbf{P}} - \left( \omega / \mu \right) \mathbf{B} = \mathbf{S} - \left( \omega / \mu \right) \mathbf{B}$.  (We are adhering to the notation employed, for instance, in Ref.~\cite{hannestad2006}.)  One can readily show that if $\omega$, $\mu$, and $\mathbf{B}$ are all constant, then Eqs.~\eqref{ppbar} lead to
\begin{align}
& \dot{\mathbf{D}} = \omega \mathbf{B} \times \mathbf{Q}, \notag \\
& \dot{\mathbf{Q}} = \mu \mathbf{D} \times \mathbf{Q}. \label{dq} 
\end{align}
It is clear that $\mathbf{D} \cdot \mathbf{Q}$ and $Q = | \mathbf{Q} |$ are both constants of motion.  Defining $\mathbf{q} = \mathbf{Q} / Q$ and $\sigma = \mathbf{D} \cdot \mathbf{Q} / Q$, Eqs.~\eqref{dq} can be used to obtain
\begin{equation}
\frac{\mathbf{q} \times \ddot{\mathbf{q}}}{\mu} = \omega Q \mathbf{B} \times \mathbf{q} - \sigma \dot{\mathbf{q}}. \label{qddot}
\end{equation}
This equation describes a gyroscopic pendulum with total angular momentum
\begin{equation}
\mathbf{D} = \frac{\mathbf{q} \times \dot{\mathbf{q}}}{\mu} + \sigma \mathbf{q},
\end{equation}
where the first term corresponds to the orbital angular momentum, the second term to the spin.  Moreover, the total energy is found to be
\begin{equation}
E = \omega Q \mathbf{B} \cdot \mathbf{q} + \frac{\mu}{2} \mathbf{D}^2. \label{energy}
\end{equation}
Interpreting the first half of the right-hand side as the potential energy, the following picture emerges: the bipolar neutrino system is equivalent to a gyroscopic pendulum, with position vector $\mathbf{Q}$, moment of inertia $\mu^{-1}$, and spin $\sigma$, swinging under the influence of a gravitational force $- \omega \mathbf{B}$ \cite{hannestad2006}.

Bipolar oscillations are thought to occur in the neutrino emission from a core-collapse supernova, during, for instance, the late-time, neutrino-driven-wind phase, when the luminosities of the individual species are comparable but there exists a stark energy hierarchy due to the differing opacities of these species in the outflowing material: $\left\langle E_{\nu_e} \right\rangle < \left\langle E_{\bar{\nu}_e} \right\rangle < \left\langle E_{\nu_\beta} \right\rangle \approx \left\langle E_{\bar{\nu}_\beta} \right\rangle$, $\beta = \mu, \tau$.  This scenario, in fact, has been the standard one in studies of the gyroscopic pendulum

The above hierarchy, however, is not universally applicable, and we find, by considering other arrangements, that the gyroscopic pendulum can exhibit rather bizarre precession behavior.  As an illustration, we consider neutrino flavor transformation in the early universe in the presence of a nonzero lepton number.  Lepton asymmetries are not only weakly constrained by present data \cite{mangano2012, steigman2012, castorina2012, barenboim2017} but, in the case of lepton asymmetries much larger than the baryon asymmetry, are motivated by leptogenesis models \cite{shaposhnikov2008, eijima2017} and are integral to a viable production mechanism for sterile neutrino dark matter \cite{shi1999, abazajian2001b, kishimoto2008}.

Prior to any significant flavor conversion, neutrinos are described by Fermi--Dirac equilibrium distribution functions, with the number density of flavor $\beta$ given by $n_{\nu_\beta} = \left( T^3 / 2 \pi^2 \right) F_2 \left( \eta_\beta \right)$, where $\eta_\beta = \mu_\beta / T$ is the degeneracy parameter, defined in terms of the chemical potential $\mu_\beta$, and $F_2 \left( \eta_\beta \right) = \int_0^\infty dx \frac{x^2}{e^{x - \eta_\beta} + 1}$ is the relativistic Fermi integral.  We work with two-flavor mixing between $\nu_e$ and a state $\nu_x$ representing some particular superposition of $\nu_\mu$ and $\nu_\tau$.  Since chemical equilibrium obtains at high temperature, antineutrinos of flavor $\beta$ have degeneracy $- \eta_\beta$.

In the bipolar regime, absent a matter background, the flavor evolution of the system is dictated by Eqs.~\eqref{ppbar}, with the asymmetry between $| \mathbf{P} |$ and $| \bar{ \mathbf{P} } |$ parameterized by
\begin{equation}
\alpha = \frac{\bar{P}_z (T_i)}{P_z (T_i)} = \frac{F_2 (- \eta_e) - F_2(- \eta_x)}{F_2 (+ \eta_e) - F_2( + \eta_x)},
\end{equation}
where the system is taken to be comprised of flavor eigenstates at initial temperature $T_i$.  To first order in the degeneracy parameters, $\alpha \approx -1 + \left( 12 / \pi^2 \right) \log 2 \left( \eta_e + \eta_x \right)$, indicating that $\mathbf{P}$ and $\bar{\mathbf{P}}$ are \textit{antialigned} at high temperature, in contrast to the initial alignment of the polarization vectors that is typical of supernova neutrino fluxes.  Moreover, by taking the $T \longrightarrow \infty$ limit and dropping prefactors roughly of order unity, the position and angular-momentum vectors are found to have magnitudes $Q \sim | \eta_e^2 - \eta_x^2 |$ and $D = | \mathbf{D} | \sim | \eta_e - \eta_x |$ (see Supplemental Material): the angular momentum is parametrically enhanced relative to the length of the pendulum.  The significance of this point will appear shortly. 

\begin{figure*}
\centering
\begin{subfigure}{0.47\textwidth}
\includegraphics[width=\textwidth]{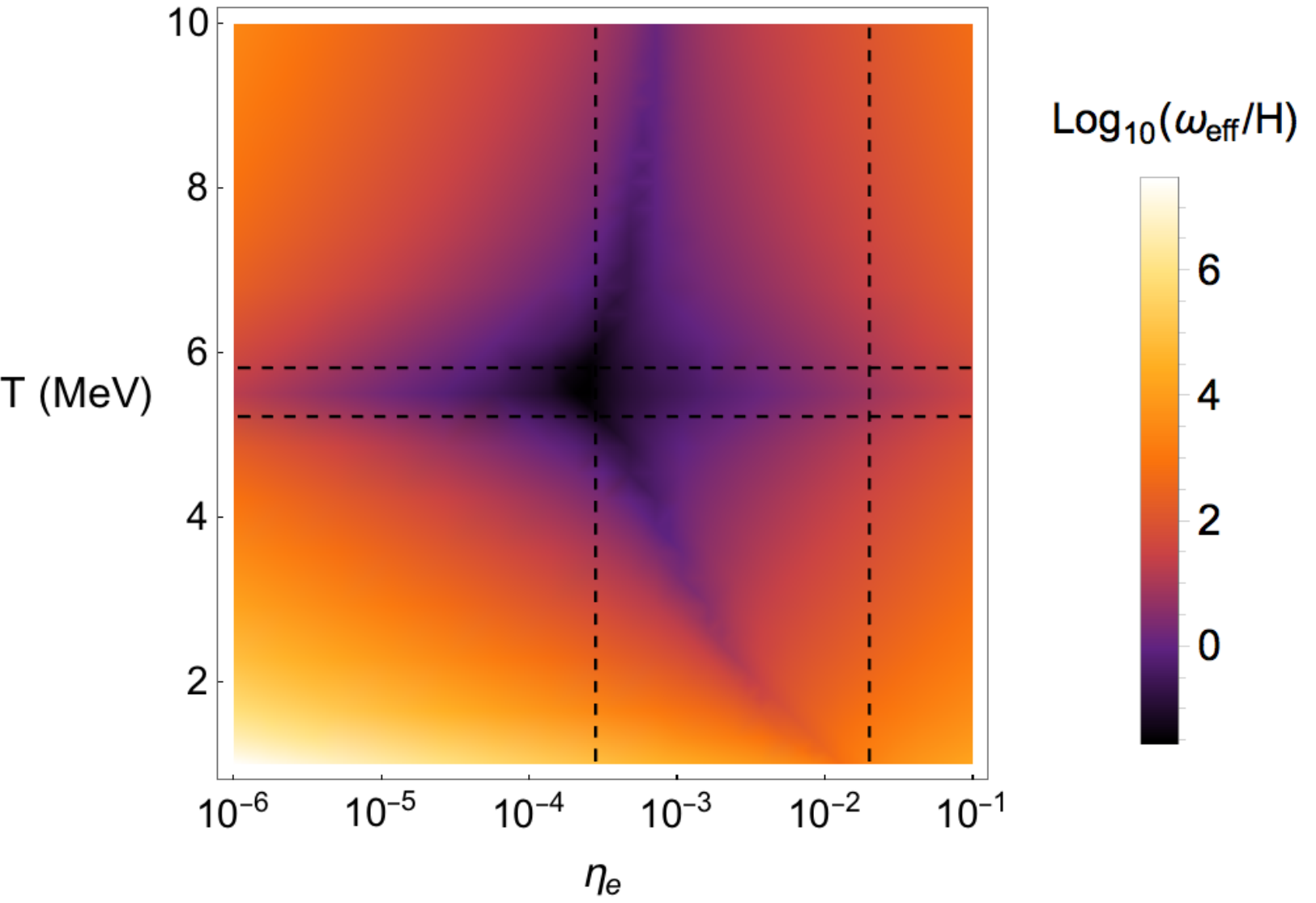}
\end{subfigure} ~~
\begin{subfigure}{0.47\textwidth}
\includegraphics[width=\textwidth]{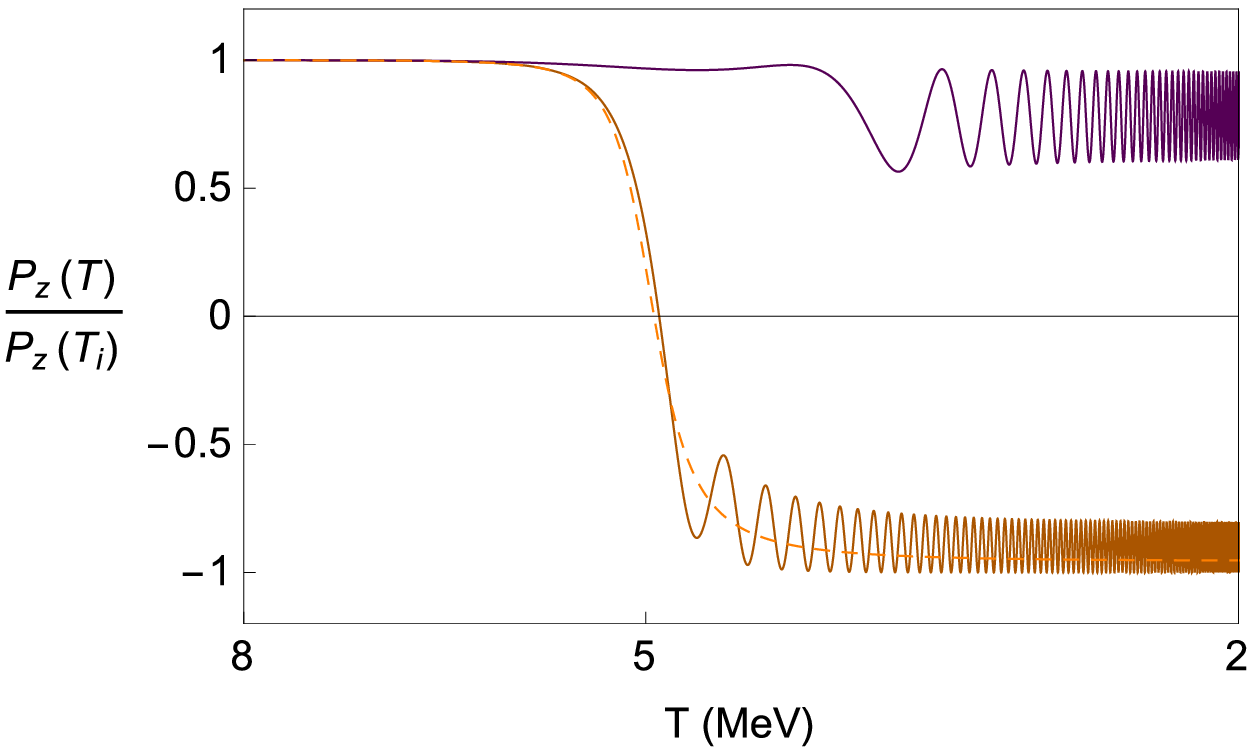}
\end{subfigure}
\caption{(Left) Ratio of $\omega_\textrm{eff}$ to the Hubble rate $H$ as a function of temperature $T$ and chemical potential $\eta_e$, calculated using pre-oscillation flavor states and the appropriate modification of Eq.~\eqref{weff} needed to accommodate a matter background (as detailed in the Supplemental Material).  The dashed horizontal lines bracket the matter-only MSW resonance width; the vertical lines correspond to the two solid curves in the plot to the right.  (Right) $P_z (T) / P_z (T_i) = \left[ n_{\nu_e}(T) - n_{\nu_x}(T) \right] / \left[ n_{\nu_e}(T_i) - n_{\nu_x}(T_i) \right]$, for $\eta_x = 0$ and $\eta_e = 10^{-8}$ (dashed orange), $2.8\times10^{-4}$ (solid purple), and $2\times10^{-2}$ (solid burnt orange).  The first of these is indistinguishable from matter-only MSW conversion, due to the small asymmetry.}
\label{densitypz}
\end{figure*}

In the early universe Eqs.~\eqref{dq} are slightly modified: while the equation of motion for $\mathbf{D}$ is unchanged, the second line becomes $\dot{\mathbf{Q}} = \mu \mathbf{D} \times \mathbf{Q} - 4 H \left( \omega / \mu \right) \mathbf{B}$, where $H$ is the Hubble constant.  In principle the expansion of the universe, which induces time-dependence in $\omega$ and $\mu$, causes the behavior of the system to be quite complicated, but analytical insights may be gained by making a few observations.  Firstly, there are two quantities that are strictly conserved in spite of the redshift:
\begin{equation}
\mathbf{B} \cdot \mathbf{D} = \textrm{constant}, \label{const1}
\end{equation}
which can be interpreted as the magnitude (up to sign) of the angular momentum along the gravitational field, and
\begin{equation}
\mathbf{D} \cdot \mathbf{Q} + \frac{\omega}{\mu} \mathbf{B} \cdot \mathbf{D} = \textrm{constant}, \label{const2}
\end{equation}
which ultimately encodes unitarity in the flavor evolution.  Furthermore, although the ``total energy'' in Eq.~\eqref{energy} is not truly conserved, the gyroscopic-pendulum picture is still valid over time scales shorter than a Hubble time.  In particular, the pendulum in the early universe may be regarded as being dominated by kinetic energy, in the sense that the second term in Eq.~\eqref{energy} is much larger than the first one all the way down to very low temperature, when the neutrino number density finally becomes sufficiently dilute.  This claim follows from the observation that, by the estimates above, $Q$ is roughly of the same order as $\mathbf{D}^2$ and therefore the ratio of potential to kinetic energy is $\sim \omega / \mu$.  The pendulum can only transfer a limited fraction of its total energy to potential energy (\textit{i.e.}, by standing straight up against gravity).  Since the kinetic energy vastly exceeds this maximum potential energy, the system is prevented from significantly draining its kinetic energy.  As a consequence, $D$ is roughly constant.

Since both $D$ and $\mathbf{B} \cdot \mathbf{D}$ are (approximately) constant, the angular momentum is well described by
\begin{equation}
\dot{\mathbf{D}} \cong \omega_\textrm{eff} \mathbf{B} \times \mathbf{D}, \label{deff}
\end{equation}
where $\omega_\textrm{eff}$ is the frequency that emerges because $\mathbf{D}$ is effectively constrained to precess about $\mathbf{B}$.  Since the first line of Eq.~\eqref{dq} must still be satisfied, this frequency is deduced to be
\begin{equation}
\omega_\textrm{eff} = \omega \frac{Q \sigma - \left( \mathbf{B} \cdot \mathbf{Q} \right) \left( \mathbf{B} \cdot \mathbf{D} \right)}{\mathbf{D}^2 - \left( \mathbf{B} \cdot \mathbf{D} \right)^2}, \label{weff}
\end{equation}
using $\mathbf{D} \cdot \mathbf{Q} = Q \sigma$. Employing Eqs.~\eqref{const1} and \eqref{const2} and $D \sim \textrm{constant}$, one finds that the only impediment to calculating $\omega_\textrm{eff}$ ``by hand'' is the presence of the factor $\mathbf{B} \cdot \mathbf{Q}$.

Even with this a priori unknown factor we can draw important qualitative conclusions from Eqs.~\eqref{deff} and \eqref{weff}.  It can be shown that, in the normal mass hierarchy, the spin $\sigma$ \textit{at high temperature} has the same sign as $\mathbf{B} \cdot \mathbf{D}$ \textit{at low temperature} if and only if $\eta_e^2 < \eta_x^2$; in the inverted hierarchy, the condition is $\eta_e^2 > \eta_x^2$.  If the signs match, then Eq.~\eqref{const2} requires $\sigma$ to reverse its sign at some point as the universe cools.  Put another way, the blueshifting of $\omega / \mu$ causes the spin of the gyroscope to slow down to a stop and then to spin up in the opposite direction.

In the course of $\sigma$ changing sign, $\omega_\textrm{eff}$ itself goes through zero.  When this occurs, $\mathbf{B} \times \mathbf{Q} \cong 0$: the gyroscopic pendulum momentarily stands upright against gravity.  But recalling the definition $\mathbf{Q} = \mathbf{S} - \left( \omega / \mu \right) \mathbf{B}$, one discerns that the pendulum must subsequently fall to a lower height as the growing factor $\omega / \mu$ progressively weighs it down.  In falling from an inverted to a normal pendulum, the gyroscope continues its precession---but now with the opposite handedness.  This qualitative analysis has been verified numerically and is visualized in Fig.~\ref{top}.

More significant from the perspective of flavor transformation is the behavior of the angular-momentum vector during this period.  As $\omega_\textrm{eff}$ passes through zero, $\mathbf{D}$ comes nearly to a dead stop and then begins, like $\mathbf{Q}$, to precess with the opposite handedness.  The importance of this is that $\mathbf{D} (t) \approx 2 \mathbf{P} (t) \approx -2 \bar{\mathbf{P}} (t)$, on account of the initial condition $\bar{\mathbf{P}} (t_i) \approx - \mathbf{P} (t_i)$ and the nature of Eqs.~\eqref{ppbar}.  The evolution of $\mathbf{D}$ therefore reflects the transformation of flavor: when the precession of $\mathbf{D}$ comes to a halt and then reverses, so too do the oscillations of $\mathbf{P}$ and $\bar{\mathbf{P}}$.

Amusingly, this mechanism can be analogized, with some poetic license, to the behavior of a real-life, canoe-shaped toy known as a rattleback.  When spun in one direction, a rattleback rattles to a halt and then spins back in the opposite direction.  The rattling is caused by the growth of a rotational instability due to the misalignment of the principal axes of the toy and the symmetry axis of its bottom surface \cite{case2014}.  The gyroscopic pendulum does not rattle, but it too has a sort of instability that causes a reversal: in this case it is the misalignment of the neutrino mass and flavor axes that facilitates the growth of the instability.

Up to this point we have ignored the electrons and positrons in the background, but the lessons from the foregoing analysis carry over straightforwardly, for the following reasons.  At some temperature, which we will denote by $T_\textrm{MSW}$, the refractive contribution from the $e^\pm$ background transitions from dominant ($T \gtrsim T_\textrm{MSW}$) to subdominant ($T \lesssim T_\textrm{MSW}$) relative to the contribution from neutrino mass.  At $T \gtrsim T_\textrm{MSW}$, the principal effect of $e^\pm$ is to suppress flavor mixing, while at $T \lesssim T_\textrm{MSW}$ the effects are negligible.  The interesting behavior is thus confined to the MSW region, $T \sim T_\textrm{MSW}$, where maximal mixing is expected to occur in the IH, at least in the absence of nonlinear neutrino--neutrino coupling.  

How does the gyroscopic pendulum evolve through the MSW region?  If the frequency $\omega_\textrm{eff}$ is fast relative to the Hubble expansion rate, then significant flavor conversion occurs as expected.  If, however, $\omega_\textrm{eff}$ happens to be close to zero at $T \sim T_\textrm{MSW}$, then conversion is stifled.  There is an intuitive visual explanation for this behavior: Imagine a rapidly precessing pendulum.  Now imagine---and this is easier said than done---rotating the orientation of gravity.  If the pendulum precesses rapidly enough, it will track the gravitational field as it rotates.  But if the precession is slow on the scale of the gravity-rotation time scale, then the pendulum will be left behind.

These considerations suggest that the precession-reversal mechanism, should it occur close to the MSW resonance, may impede flavor conversion.  It turns out, in fact, that the reversal is associated with a more general phenomenon that impairs adiabaticity even when the conditions are not met for $\sigma$ to change sign.  Inspection of Eq.~\eqref{weff} reveals that, prior to flavor transformation, the dependence of $\omega_\textrm{eff}$ on the chemical potentials factors out as $Q / D$.  Having noted earlier that $D \sim \textrm{constant}$, we now note that $Q$ is dominated, except over a small temperature range, either by the contribution from $\mathbf{S}$ or from the part proportional to $\mathbf{B}$. Using $\mu = \sqrt{2} G_F T^3$, $\omega = | \delta m^2 | / 4 \epsilon$ for comoving energy $\epsilon$, and the dependence of $\mathbf{S}$ and $\mathbf{D}$ on the chemical potentials (see Supplemental Material), the frequency scaling in these two limits is found to be $\omega_\textrm{eff} \propto \omega |\mathbf{S}| / D \propto T^{-1} | \eta_e + \eta_x |$ when $\mathbf{S}$-dominated and $\omega_\textrm{eff} \propto \omega^2 / \mu D \propto T^{-5} | \eta_e - \eta_x |^{-1}$ when $\mathbf{B}$-dominated.

The frequency at resonance is thus minimized by the smallest $| \eta_e + \eta_x |$ such that $T^{-5}$ scaling is pushed below the MSW region, or, in other words, such that the transition temperature $T_\textrm{trans}$, at which $| \mathbf{S} | \sim | - \left( \omega / \mu \right) \mathbf{B} |$, is smaller than $T_\textrm{MSW}$.  These temperatures compare as
\begin{equation}
\frac{T_\textrm{trans}}{T_\textrm{MSW}} \sim 1.913 \left( \frac{\delta m^2 ~\epsilon}{G_F m_W^4 | \eta_e^2 - \eta_x^2 |^3 \cos^2 2 \theta} \right)^{1/12}, \label{ratio}
\end{equation}
where $m_W$ is the W boson mass. For $1 - 3$ mixing parameters, this becomes $T_\textrm{trans} / T_\textrm{MSW} \sim \sqrt{ 5.87 \times 10^{-4} / \sqrt{| \eta_e^2 - \eta_x^2|} }$. Incidentally, the coefficient corresponds to a chemical potential not too far below current constraints. Note that the choice of mixing channel is virtually immaterial, owing to the small exponent.

Fig.~\ref{densitypz} shows, in a manner consistent with Eq.~\eqref{ratio}, the influence of $\omega_\textrm{eff}$ on the flavor transformation of a cosmic bipolar neutrino system. It is intriguing that, for neutrino degeneracies that are not very nearly equal in magnitude, the most profound suppression of flavor conversion occurs for lepton asymmetries in the neighborhood relevant for the resonant production of sterile neutrino dark matter \cite{shi1999, abazajian2001b, kishimoto2008, abazajian2014, bozek2016, horiuchi2016}.  

The impact of a small $\omega_\textrm{eff}$ was actually identified in an earlier study \cite{johns2016}, but the mechanism underlying the suppression of flavor conversion was then unknown.  Indeed, the estimates of the preceding paragraph seem to be borne out quite well in the numerical results of that paper.  Interpreting the results of Ref.~\cite{johns2016} in light of the present analysis, it appears that this phenomenon may be of considerable importance for neutrino flavor evolution in the early universe.  Possible connections between the suppression of flavor conversion and production of the light nuclides in Big Bang nucleosynthesis were speculated on in Ref.~\cite{johns2016} and still appear viable.

Whether the precession phenomena analyzed in this Letter manifest in compact-object environments is an open question.  A plausible candidate is the O-Ne-Mg core-collapse supernova, which is expected to occur for some progenitor stars in the mass range $\sim 8 - 10~M_\odot$.  From the viewpoint of neutrino flavor, the intriguing characteristic of this site is the extremely steep density gradient at the surface of the core, which places the MSW resonances inside the region where neutrino number density is still high and therefore collective effects are still influential \cite{duan2008b, dasgupta2008c, lunardini2008, cherry2010}.  Sufficiently adiabatic MSW conversion may conceivably manufacture a hierarchy of fluxes such that the foregoing analysis, with proper modifications, is applicable post-resonance.  In a similar vein, an environment in which the density of free neutrons or protons is very high outside the neutrino decoupling surface may deplete enough $\nu_e$ or $\bar{\nu}_e$ via charged-current capture to alter the initial flux hierarchy, thereby engineering the requisite conditions. Targeted numerical analysis can be used to confirm whether the foregoing speculations are borne out in these environments, where temporal and spatial instabilities and trajectory dependence \cite{duan2006, mirizzi2011} are also in play.

It is intriguing that the nonlinear problem of neutrino flavor transformation, although extensively explored, may harbor surprising phenomena---such as those pointed out in this Letter---that have implications for cosmology and compact-object physics.

\begin{acknowledgments}
This work was supported by NSF Grant Nos. PHY-1307372 and PHY-1614864 at UC San Diego.  During the completion of this work, L.J. was supported by an Inamori Fellowship.  We thank Brad Keister and Sebastien Tawa for helpful conversations.
\end{acknowledgments}

\clearpage
\begin{center}
\textbf{\large Supplemental material}
\end{center}

\section{MSW and the matter background}

In the early universe, effective in-medium mixing parameters can be introduced to account for the influence of the matter background on vacuum oscillations.  Recalling the definition $\omega = | \delta m^2 | / 4 E$ for neutrinos of energy $E$, the in-medium oscillation frequency $\omega_m$ and in-medium mixing angle $\theta_m$ are respectively given by
\begin{equation}
\omega_m = \sqrt{ \omega^2 \sin^2 2 \theta + \left( \pm \omega \cos 2\theta + \mathcal{V} \right)^2 },
\end{equation}
and
\begin{equation}
\sin^2 2 \theta_m = \frac{ \omega^2 \sin^2 2\theta}{\omega_m^2},
\end{equation}
where the $+$ ($-$) is for the normal (inverted) neutrino mass hierarchy and $\mathcal{V}$ is the potential from forward scattering of neutrinos on matter particles.  This potential is
\begin{equation}
\mathcal{V} = \frac{2 \sqrt{2} G_F E \varrho_{e^\pm}}{3 m_W^2},
\end{equation}
with Fermi coupling constant $G_F$,  $e^\pm$ energy density $\varrho_{e^\pm}$, and W boson mass $m_W$ \cite{notzold1988}.  Setting aside the impact of the neutrino self-coupling potential, an MSW resonance is expected to occur in the inverted hierarchy where $\theta_m$ is maximal.  The temperature at which this occurs is denoted $T_\textrm{MSW}$.

In a compact-object environment, the dominant contribution to the matter potential is CP-asymmetric and therefore cannot be incorporated into effective mixing parameters describing both neutrinos and antineutrinos simultaneously.  However, a co-rotating reference frame can be adopted: the matter term drops out of the Hamiltonian, but at the expense of the vacuum Hamiltonian vector $\mathbf{B}$ oscillating.  This approach is laid out in Refs.~\cite{duan2006c, hannestad2006, duan2006, duan2007b}.

\section{Nonadiabaticity and precession}

Letting $T_i$ denote a temperature above the onset of significant flavor transformation, the initial ``sum'' and ``difference'' vectors are
\begin{align}
&\mathbf{S}(T_i) = \mathbf{P}(T_i) + \bar{\mathbf{P}}(T_i) = \frac{1 + \alpha}{2 \pi^2} \left[ F_2 (\eta_e) - F_2 (\eta_x) \right] \hat{\mathbf{z}}, \notag \\
&\mathbf{D}(T_i) = \mathbf{P}(T_i) - \bar{\mathbf{P}}(T_i) = \frac{1 - \alpha}{2 \pi^2} \left[ F_2 (\eta_e) - F_2 (\eta_x) \right] \hat{\mathbf{z}}.
\end{align}
After expanding the relativistic Fermi integrals to first order, evaluating them, and using the approximation for $\alpha$ given in the Letter, the vectors become
\begin{align}
&\mathbf{S}(T_i) \approx \frac{\log 2}{\pi^2} \left( \eta_e^2 - \eta_x^2 \right) \hat{\mathbf{z}}, \notag \\
&\mathbf{D}(T_i) \approx \frac{1}{6} \left( \eta_e - \eta_x \right) \hat{\mathbf{z}}.
\end{align}
These expressions then give the scaling of $Q / D$, and thereby of $\omega_\textrm{eff}$, in the (matter-free) limits $\mathbf{Q} \sim \mathbf{S}$ and $\mathbf{Q} \sim - \left( \omega / \mu \right) \mathbf{B}$.  The transition between these two limits occurs at $T_\textrm{trans}$, found by setting $| \mathbf{S} | = | - \left( \omega / \mu \right) \mathbf{B} |$, where again the influence of matter is being neglected.  

A first approximation at including the influence of matter is to effect the replacements $\omega \longrightarrow \omega_m$ and $\theta \longrightarrow \theta_m$ in the matter-free analysis, making sure to choose the quadrant of $\theta_m$ appropriately.  For notational convenience, we also replace $\mathbf{B}$ with $\mathbf{B}_m$, differing only by the use of $\theta$ versus $\theta_m$ in the definition.  Under this prescription, the effective precession frequency of the angular-momentum vector $\mathbf{D}$ becomes
\begin{equation}
\omega_\textrm{eff} = \omega_m \frac{Q\sigma - \left( \mathbf{B}_m \cdot \mathbf{Q} \right) \left( \mathbf{B}_m \cdot \mathbf{D} \right)}{\mathbf{D}^2 - \left( \mathbf{B}_m \cdot \mathbf{D} \right)^2},
\end{equation}
where now the precession is about the (temperature-dependent) vector $\mathbf{B}_m$.  The left panel of Fig.~2 of the Letter plots this function, using pre-oscillation values of $\mathbf{Q}$ and $\mathbf{D}$.  Eq.~11 of the Letter, and the scaling limits of $\omega_\textrm{eff}$ that precede it, were calculated using the same prescription.

Absent the coupling of different neutrino modes, the adiabaticity is set by $\omega_m / H$, where $H$ is the Hubble rate. If the coupling is strong, the adiabaticity is instead set by $\omega_\textrm{eff} / H$. In the early universe, resonance is traversed adiabatically by default ($\omega_m / H \gg 1$ due to the slowness of Hubble expansion) but adiabaticity is lost if $\omega_\textrm{eff}$ is sufficiently small. As we argue in the main text, $\omega_\textrm{eff}$ scales like $T^{-1} | \eta_e + \eta_x |$ in the $\mathbf{Q} \sim \mathbf{S}$ regime ($T \gtrsim T_\textrm{trans}$) and like $T^{-5} | \eta_e - \eta_x |^{-1}$ in the $\mathbf{Q} \sim - (\omega / \mu ) \mathbf{B}$ regime ($T \lesssim T_\textrm{trans}$). The frequency at resonance is therefore minimized if the rapid increase in the latter regime is delayed at least until the resonance temperature $T_\textrm{MSW}$ is reached. On the other hand, given the scaling with neutrino degeneracy, minimizing $\omega_\textrm{eff}$ also requires minimizing $| \eta_e + \eta_x |$, such that $T_\textrm{trans} \lesssim T_\textrm{MSW}$ still holds.

\section{Non-forward scattering}

The calculations presented in the Letter were performed neglecting non-forward scattering (\textit{i.e.}, collisions) \cite{sigl1993, strack2005, vlasenko2014, serreau2014, blaschke2016}. In Fig.~1 we display the results of repeating the calculations shown in the right panel of Fig.~2 of the main text but with collisions now accounted for. The suppression of flavor conversion witnessed in the collisionless analysis translates to a suppression of flavor equilibration $\left( P_z \rightarrow 0 \right)$ in the calculation with collisions.

In producing Fig.~1 we have employed the damping approximation used in Refs.~\cite{stodolsky1987, thomson1992, bell1999, dolgov2002} and elsewhere. With this prescription, the equations of motion, given in Eq.~1 of the main text, become
\begin{align}
& \dot{\mathbf{P}} = \left( +\omega \mathbf{B} - \mu \bar{\mathbf{P}} \right) \times \mathbf{P} - D \mathbf{P}_T, \notag \\
& \dot{\bar{\mathbf{P}}} = \left( -\omega \mathbf{B} + \mu \mathbf{P} \right) \times \bar{\mathbf{P}} - D \bar{\mathbf{P}}_T,
\end{align}
where the subscript $T$ denotes the part of the vector transverse to the flavor ($z$) axis and $D$ is a damping coefficient proportional to the scattering rate. The expression for $D$ and the derivation of this prescription can be found in the references above.

By a plasma temperature of 1 MeV neutrinos have begun to decouple and subsequent effects of collisions are small. The survival of coherent (collisionless) flavor phenomena through to the epochs of neutrino decoupling and Big Bang nucleosynthesis is discussed at greater length in Ref.~\cite{johns2016}.

\begin{figure}
\includegraphics[width=.45\textwidth]{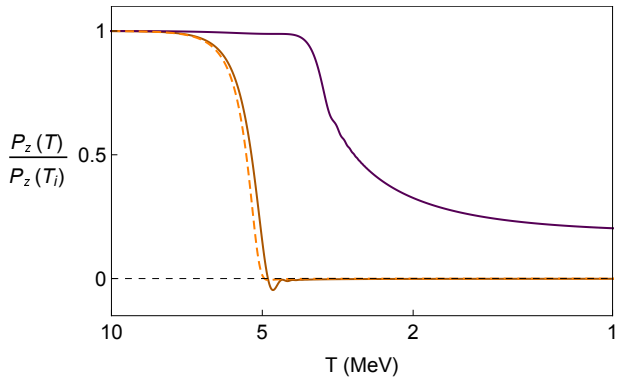}
\caption{Identical to the right panel of Fig. 2 of the main text except that the influence of collisions is accounted for here. The curves correspond to lepton asymmetries of $\eta_x = 0$ and $\eta_e = 10^{-8}$ (dashed orange), $2.8\times10^{-4}$ (solid purple), and $2\times10^{-2}$ (solid burnt orange).}  
\label{damped}
\end{figure}

\clearpage
\bibliography{all_papers}

\begin{thebibliography}{92}%
\makeatletter
\providecommand \@ifxundefined [1]{%
 \@ifx{#1\undefined}
}%
\providecommand \@ifnum [1]{%
 \ifnum #1\expandafter \@firstoftwo
 \else \expandafter \@secondoftwo
 \fi
}%
\providecommand \@ifx [1]{%
 \ifx #1\expandafter \@firstoftwo
 \else \expandafter \@secondoftwo
 \fi
}%
\providecommand \natexlab [1]{#1}%
\providecommand \enquote  [1]{``#1''}%
\providecommand \bibnamefont  [1]{#1}%
\providecommand \bibfnamefont [1]{#1}%
\providecommand \citenamefont [1]{#1}%
\providecommand \href@noop [0]{\@secondoftwo}%
\providecommand \href [0]{\begingroup \@sanitize@url \@href}%
\providecommand \@href[1]{\@@startlink{#1}\@@href}%
\providecommand \@@href[1]{\endgroup#1\@@endlink}%
\providecommand \@sanitize@url [0]{\catcode `\\12\catcode `\$12\catcode
  `\&12\catcode `\#12\catcode `\^12\catcode `\_12\catcode `\%12\relax}%
\providecommand \@@startlink[1]{}%
\providecommand \@@endlink[0]{}%
\providecommand \url  [0]{\begingroup\@sanitize@url \@url }%
\providecommand \@url [1]{\endgroup\@href {#1}{\urlprefix }}%
\providecommand \urlprefix  [0]{URL }%
\providecommand \Eprint [0]{\href }%
\providecommand \doibase [0]{http://dx.doi.org/}%
\providecommand \selectlanguage [0]{\@gobble}%
\providecommand \bibinfo  [0]{\@secondoftwo}%
\providecommand \bibfield  [0]{\@secondoftwo}%
\providecommand \translation [1]{[#1]}%
\providecommand \BibitemOpen [0]{}%
\providecommand \bibitemStop [0]{}%
\providecommand \bibitemNoStop [0]{.\EOS\space}%
\providecommand \EOS [0]{\spacefactor3000\relax}%
\providecommand \BibitemShut  [1]{\csname bibitem#1\endcsname}%
\let\auto@bib@innerbib\@empty
\bibitem [{\citenamefont {Shi}\ and\ \citenamefont {Fuller}(1999)}]{shi1999}%
  \BibitemOpen
  \bibfield  {author} {\bibinfo {author} {\bibfnamefont {X.}~\bibnamefont
  {Shi}}\ and\ \bibinfo {author} {\bibfnamefont {G.~M.}\ \bibnamefont
  {Fuller}},\ }\href@noop {} {\bibfield  {journal} {\bibinfo  {journal} {Phys.
  Rev. Lett.}\ }\textbf {\bibinfo {volume} {82}},\ \bibinfo {pages} {2832}
  (\bibinfo {year} {1999})}\BibitemShut {NoStop}%
\bibitem [{\citenamefont {Abazajian}\ \emph {et~al.}(2001)\citenamefont
  {Abazajian}, \citenamefont {Fuller},\ and\ \citenamefont
  {Patel}}]{abazajian2001b}%
  \BibitemOpen
  \bibfield  {author} {\bibinfo {author} {\bibfnamefont {K.}~\bibnamefont
  {Abazajian}}, \bibinfo {author} {\bibfnamefont {G.~M.}\ \bibnamefont
  {Fuller}}, \ and\ \bibinfo {author} {\bibfnamefont {M.}~\bibnamefont
  {Patel}},\ }\href@noop {} {\bibfield  {journal} {\bibinfo  {journal} {Phys.
  Rev. D}\ }\textbf {\bibinfo {volume} {64}},\ \bibinfo {pages} {023501}
  (\bibinfo {year} {2001})}\BibitemShut {NoStop}%
\bibitem [{\citenamefont {Asaka}\ and\ \citenamefont
  {Shaposhnikov}(2005)}]{asaka2005}%
  \BibitemOpen
  \bibfield  {author} {\bibinfo {author} {\bibfnamefont {T.}~\bibnamefont
  {Asaka}}\ and\ \bibinfo {author} {\bibfnamefont {M.}~\bibnamefont
  {Shaposhnikov}},\ }\href {\doibase
  http://dx.doi.org/10.1016/j.physletb.2005.06.020} {\bibfield  {journal}
  {\bibinfo  {journal} {Phys. Lett. B}\ }\textbf {\bibinfo {volume} {620}},\
  \bibinfo {pages} {17 } (\bibinfo {year} {2005})}\BibitemShut {NoStop}%
\bibitem [{\citenamefont {Asaka}\ \emph {et~al.}(2005)\citenamefont {Asaka},
  \citenamefont {Blanchet},\ and\ \citenamefont {Shaposhnikov}}]{asaka2005b}%
  \BibitemOpen
  \bibfield  {author} {\bibinfo {author} {\bibfnamefont {T.}~\bibnamefont
  {Asaka}}, \bibinfo {author} {\bibfnamefont {S.}~\bibnamefont {Blanchet}}, \
  and\ \bibinfo {author} {\bibfnamefont {M.}~\bibnamefont {Shaposhnikov}},\
  }\href {\doibase http://dx.doi.org/10.1016/j.physletb.2005.09.070} {\bibfield
   {journal} {\bibinfo  {journal} {Phys. Lett. B}\ }\textbf {\bibinfo {volume}
  {631}},\ \bibinfo {pages} {151 } (\bibinfo {year} {2005})}\BibitemShut
  {NoStop}%
\bibitem [{\citenamefont {Kishimoto}\ and\ \citenamefont
  {Fuller}(2008)}]{kishimoto2008}%
  \BibitemOpen
  \bibfield  {author} {\bibinfo {author} {\bibfnamefont {C.~T.}\ \bibnamefont
  {Kishimoto}}\ and\ \bibinfo {author} {\bibfnamefont {G.~M.}\ \bibnamefont
  {Fuller}},\ }\href@noop {} {\bibfield  {journal} {\bibinfo  {journal} {Phys.
  Rev. D}\ }\textbf {\bibinfo {volume} {78}},\ \bibinfo {pages} {023524}
  (\bibinfo {year} {2008})}\BibitemShut {NoStop}%
\bibitem [{\citenamefont {Petraki}\ and\ \citenamefont
  {Kusenko}(2008)}]{petraki2008}%
  \BibitemOpen
  \bibfield  {author} {\bibinfo {author} {\bibfnamefont {K.}~\bibnamefont
  {Petraki}}\ and\ \bibinfo {author} {\bibfnamefont {A.}~\bibnamefont
  {Kusenko}},\ }\href {\doibase 10.1103/PhysRevD.77.065014} {\bibfield
  {journal} {\bibinfo  {journal} {Phys. Rev. D}\ }\textbf {\bibinfo {volume}
  {77}},\ \bibinfo {pages} {065014} (\bibinfo {year} {2008})}\BibitemShut
  {NoStop}%
\bibitem [{\citenamefont {Bezrukov}\ \emph {et~al.}(2013)\citenamefont
  {Bezrukov}, \citenamefont {Kartavtsev},\ and\ \citenamefont
  {Lindner}}]{bezrukov2013}%
  \BibitemOpen
  \bibfield  {author} {\bibinfo {author} {\bibfnamefont {F.}~\bibnamefont
  {Bezrukov}}, \bibinfo {author} {\bibfnamefont {A.}~\bibnamefont
  {Kartavtsev}}, \ and\ \bibinfo {author} {\bibfnamefont {M.}~\bibnamefont
  {Lindner}},\ }\href {http://stacks.iop.org/0954-3899/40/i=9/a=095202}
  {\bibfield  {journal} {\bibinfo  {journal} {J. Phys. G}\ }\textbf {\bibinfo
  {volume} {40}},\ \bibinfo {pages} {095202} (\bibinfo {year}
  {2013})}\BibitemShut {NoStop}%
\bibitem [{\citenamefont {Venumadhav}\ \emph {et~al.}(2016)\citenamefont
  {Venumadhav}, \citenamefont {Cyr-Racine}, \citenamefont {Abazajian},\ and\
  \citenamefont {Hirata}}]{venumadhav2016}%
  \BibitemOpen
  \bibfield  {author} {\bibinfo {author} {\bibfnamefont {T.}~\bibnamefont
  {Venumadhav}}, \bibinfo {author} {\bibfnamefont {F.-Y.}\ \bibnamefont
  {Cyr-Racine}}, \bibinfo {author} {\bibfnamefont {K.~N.}\ \bibnamefont
  {Abazajian}}, \ and\ \bibinfo {author} {\bibfnamefont {C.~M.}\ \bibnamefont
  {Hirata}},\ }\href {\doibase 10.1103/PhysRevD.94.043515} {\bibfield
  {journal} {\bibinfo  {journal} {Phys. Rev. D}\ }\textbf {\bibinfo {volume}
  {94}},\ \bibinfo {pages} {043515} (\bibinfo {year} {2016})}\BibitemShut
  {NoStop}%
\bibitem [{\citenamefont {Adhikari}\ \emph {et~al.}(2017)\citenamefont
  {Adhikari}, \citenamefont {Agostini}, \citenamefont {Ky}, \citenamefont
  {Araki}, \citenamefont {Archidiacono} \emph {et~al.}}]{adhikari2017}%
  \BibitemOpen
  \bibfield  {author} {\bibinfo {author} {\bibfnamefont {R.}~\bibnamefont
  {Adhikari}}, \bibinfo {author} {\bibfnamefont {M.}~\bibnamefont {Agostini}},
  \bibinfo {author} {\bibfnamefont {N.~A.}\ \bibnamefont {Ky}}, \bibinfo
  {author} {\bibfnamefont {T.}~\bibnamefont {Araki}}, \bibinfo {author}
  {\bibfnamefont {M.}~\bibnamefont {Archidiacono}},  \emph {et~al.},\ }\href
  {http://stacks.iop.org/1475-7516/2017/i=01/a=025} {\bibfield  {journal}
  {\bibinfo  {journal} {J. Cosmol. Astropart. Phys.}\ }\textbf {\bibinfo
  {volume} {2017}},\ \bibinfo {pages} {025} (\bibinfo {year}
  {2017})}\BibitemShut {NoStop}%
\bibitem [{\citenamefont {Akhmedov}\ \emph {et~al.}(1998)\citenamefont
  {Akhmedov}, \citenamefont {Rubakov},\ and\ \citenamefont
  {Smirnov}}]{akhmedov1998}%
  \BibitemOpen
  \bibfield  {author} {\bibinfo {author} {\bibfnamefont {E.~K.}\ \bibnamefont
  {Akhmedov}}, \bibinfo {author} {\bibfnamefont {V.~A.}\ \bibnamefont
  {Rubakov}}, \ and\ \bibinfo {author} {\bibfnamefont {A.~Y.}\ \bibnamefont
  {Smirnov}},\ }\href {\doibase 10.1103/PhysRevLett.81.1359} {\bibfield
  {journal} {\bibinfo  {journal} {Phys. Rev. Lett.}\ }\textbf {\bibinfo
  {volume} {81}},\ \bibinfo {pages} {1359} (\bibinfo {year}
  {1998})}\BibitemShut {NoStop}%
\bibitem [{\citenamefont {Abada}\ \emph {et~al.}(2006)\citenamefont {Abada},
  \citenamefont {Davidson}, \citenamefont {Josse-Michaux}, \citenamefont
  {Losada},\ and\ \citenamefont {Riotto}}]{abada2006}%
  \BibitemOpen
  \bibfield  {author} {\bibinfo {author} {\bibfnamefont {A.}~\bibnamefont
  {Abada}}, \bibinfo {author} {\bibfnamefont {S.}~\bibnamefont {Davidson}},
  \bibinfo {author} {\bibfnamefont {F.-X.}\ \bibnamefont {Josse-Michaux}},
  \bibinfo {author} {\bibfnamefont {M.}~\bibnamefont {Losada}}, \ and\ \bibinfo
  {author} {\bibfnamefont {A.}~\bibnamefont {Riotto}},\ }\href
  {http://stacks.iop.org/1475-7516/2006/i=04/a=004} {\bibfield  {journal}
  {\bibinfo  {journal} {J. Cosmol. Astropart. Phys.}\ }\textbf {\bibinfo
  {volume} {2006}},\ \bibinfo {pages} {004} (\bibinfo {year}
  {2006})}\BibitemShut {NoStop}%
\bibitem [{\citenamefont {Nardi}\ \emph {et~al.}(2006)\citenamefont {Nardi},
  \citenamefont {Nir}, \citenamefont {Roulet},\ and\ \citenamefont
  {Racker}}]{nardi2006}%
  \BibitemOpen
  \bibfield  {author} {\bibinfo {author} {\bibfnamefont {E.}~\bibnamefont
  {Nardi}}, \bibinfo {author} {\bibfnamefont {Y.}~\bibnamefont {Nir}}, \bibinfo
  {author} {\bibfnamefont {E.}~\bibnamefont {Roulet}}, \ and\ \bibinfo {author}
  {\bibfnamefont {J.}~\bibnamefont {Racker}},\ }\href
  {http://stacks.iop.org/1126-6708/2006/i=01/a=164} {\bibfield  {journal}
  {\bibinfo  {journal} {J. High Energy Phys.}\ }\textbf {\bibinfo {volume}
  {2006}},\ \bibinfo {pages} {164} (\bibinfo {year} {2006})}\BibitemShut
  {NoStop}%
\bibitem [{\citenamefont {Blanchet}\ and\ \citenamefont
  {Bari}(2007)}]{blanchet2007}%
  \BibitemOpen
  \bibfield  {author} {\bibinfo {author} {\bibfnamefont {S.}~\bibnamefont
  {Blanchet}}\ and\ \bibinfo {author} {\bibfnamefont {P.~D.}\ \bibnamefont
  {Bari}},\ }\href {http://stacks.iop.org/1475-7516/2007/i=03/a=018} {\bibfield
   {journal} {\bibinfo  {journal} {J. Cosmol. Astropart. Phys.}\ }\textbf
  {\bibinfo {volume} {2007}},\ \bibinfo {pages} {018} (\bibinfo {year}
  {2007})}\BibitemShut {NoStop}%
\bibitem [{\citenamefont {Shaposhnikov}(2008)}]{shaposhnikov2008}%
  \BibitemOpen
  \bibfield  {author} {\bibinfo {author} {\bibfnamefont {M.}~\bibnamefont
  {Shaposhnikov}},\ }\href {http://stacks.iop.org/1126-6708/2008/i=08/a=008}
  {\bibfield  {journal} {\bibinfo  {journal} {J. High Energy Phys.}\ }\textbf
  {\bibinfo {volume} {2008}},\ \bibinfo {pages} {008} (\bibinfo {year}
  {2008})}\BibitemShut {NoStop}%
\bibitem [{\citenamefont {Canetti}\ \emph {et~al.}(2013)\citenamefont
  {Canetti}, \citenamefont {Drewes}, \citenamefont {Frossard},\ and\
  \citenamefont {Shaposhnikov}}]{canetti2013}%
  \BibitemOpen
  \bibfield  {author} {\bibinfo {author} {\bibfnamefont {L.}~\bibnamefont
  {Canetti}}, \bibinfo {author} {\bibfnamefont {M.}~\bibnamefont {Drewes}},
  \bibinfo {author} {\bibfnamefont {T.}~\bibnamefont {Frossard}}, \ and\
  \bibinfo {author} {\bibfnamefont {M.}~\bibnamefont {Shaposhnikov}},\ }\href
  {\doibase 10.1103/PhysRevD.87.093006} {\bibfield  {journal} {\bibinfo
  {journal} {Phys. Rev. D}\ }\textbf {\bibinfo {volume} {87}},\ \bibinfo
  {pages} {093006} (\bibinfo {year} {2013})}\BibitemShut {NoStop}%
\bibitem [{\citenamefont {Dolgov}\ \emph {et~al.}(2002)\citenamefont {Dolgov},
  \citenamefont {Hansen}, \citenamefont {Pastor}, \citenamefont {Petcov},
  \citenamefont {Raffelt},\ and\ \citenamefont {Semikoz}}]{dolgov2002}%
  \BibitemOpen
  \bibfield  {author} {\bibinfo {author} {\bibfnamefont {A.~D.}\ \bibnamefont
  {Dolgov}}, \bibinfo {author} {\bibfnamefont {S.~H.}\ \bibnamefont {Hansen}},
  \bibinfo {author} {\bibfnamefont {S.}~\bibnamefont {Pastor}}, \bibinfo
  {author} {\bibfnamefont {S.~T.}\ \bibnamefont {Petcov}}, \bibinfo {author}
  {\bibfnamefont {G.~G.}\ \bibnamefont {Raffelt}}, \ and\ \bibinfo {author}
  {\bibfnamefont {D.~V.}\ \bibnamefont {Semikoz}},\ }\href@noop {} {\bibfield
  {journal} {\bibinfo  {journal} {Nucl. Phys.}\ }\textbf {\bibinfo {volume}
  {B632}},\ \bibinfo {pages} {363} (\bibinfo {year} {2002})}\BibitemShut
  {NoStop}%
\bibitem [{\citenamefont {Abazajian}\ \emph {et~al.}(2002)\citenamefont
  {Abazajian}, \citenamefont {Beacom},\ and\ \citenamefont
  {Bell}}]{abazajian2002}%
  \BibitemOpen
  \bibfield  {author} {\bibinfo {author} {\bibfnamefont {K.~N.}\ \bibnamefont
  {Abazajian}}, \bibinfo {author} {\bibfnamefont {J.~F.}\ \bibnamefont
  {Beacom}}, \ and\ \bibinfo {author} {\bibfnamefont {N.~F.}\ \bibnamefont
  {Bell}},\ }\href {\doibase 10.1103/PhysRevD.66.013008} {\bibfield  {journal}
  {\bibinfo  {journal} {Phys. Rev. D}\ }\textbf {\bibinfo {volume} {66}},\
  \bibinfo {pages} {013008} (\bibinfo {year} {2002})}\BibitemShut {NoStop}%
\bibitem [{\citenamefont {Wong}(2002)}]{wong2002}%
  \BibitemOpen
  \bibfield  {author} {\bibinfo {author} {\bibfnamefont {Y.~Y.~Y.}\
  \bibnamefont {Wong}},\ }\href {\doibase 10.1103/PhysRevD.66.025015}
  {\bibfield  {journal} {\bibinfo  {journal} {Phys. Rev. D}\ }\textbf {\bibinfo
  {volume} {66}},\ \bibinfo {pages} {025015} (\bibinfo {year}
  {2002})}\BibitemShut {NoStop}%
\bibitem [{\citenamefont {Simha}\ and\ \citenamefont
  {Steigman}(2008)}]{simha2008}%
  \BibitemOpen
  \bibfield  {author} {\bibinfo {author} {\bibfnamefont {V.}~\bibnamefont
  {Simha}}\ and\ \bibinfo {author} {\bibfnamefont {G.}~\bibnamefont
  {Steigman}},\ }\href {http://stacks.iop.org/1475-7516/2008/i=08/a=011}
  {\bibfield  {journal} {\bibinfo  {journal} {J. Cosmol. Astropart. Phys.}\
  }\textbf {\bibinfo {volume} {2008}},\ \bibinfo {pages} {011} (\bibinfo {year}
  {2008})}\BibitemShut {NoStop}%
\bibitem [{\citenamefont {Mangano}\ \emph {et~al.}(2011)\citenamefont
  {Mangano}, \citenamefont {Miele}, \citenamefont {Pastor}, \citenamefont
  {Pisanti},\ and\ \citenamefont {Sarikas}}]{mangano2011}%
  \BibitemOpen
  \bibfield  {author} {\bibinfo {author} {\bibfnamefont {G.}~\bibnamefont
  {Mangano}}, \bibinfo {author} {\bibfnamefont {G.}~\bibnamefont {Miele}},
  \bibinfo {author} {\bibfnamefont {S.}~\bibnamefont {Pastor}}, \bibinfo
  {author} {\bibfnamefont {O.}~\bibnamefont {Pisanti}}, \ and\ \bibinfo
  {author} {\bibfnamefont {S.}~\bibnamefont {Sarikas}},\ }\href
  {http://stacks.iop.org/1475-7516/2011/i=03/a=035} {\bibfield  {journal}
  {\bibinfo  {journal} {J. Cosmol. Astropart. Phys.}\ }\textbf {\bibinfo
  {volume} {2011}},\ \bibinfo {pages} {035} (\bibinfo {year}
  {2011})}\BibitemShut {NoStop}%
\bibitem [{\citenamefont {Mangano}\ \emph {et~al.}(2012)\citenamefont
  {Mangano}, \citenamefont {Miele}, \citenamefont {Pastor}, \citenamefont
  {Pisanti},\ and\ \citenamefont {Sarikas}}]{mangano2012}%
  \BibitemOpen
  \bibfield  {author} {\bibinfo {author} {\bibfnamefont {G.}~\bibnamefont
  {Mangano}}, \bibinfo {author} {\bibfnamefont {G.}~\bibnamefont {Miele}},
  \bibinfo {author} {\bibfnamefont {S.}~\bibnamefont {Pastor}}, \bibinfo
  {author} {\bibfnamefont {O.}~\bibnamefont {Pisanti}}, \ and\ \bibinfo
  {author} {\bibfnamefont {S.}~\bibnamefont {Sarikas}},\ }\href {\doibase
  http://dx.doi.org/10.1016/j.physletb.2012.01.015} {\bibfield  {journal}
  {\bibinfo  {journal} {Phys. Lett. B}\ }\textbf {\bibinfo {volume} {708}},\
  \bibinfo {pages} {1 } (\bibinfo {year} {2012})}\BibitemShut {NoStop}%
\bibitem [{\citenamefont {Pastor}\ \emph {et~al.}(2009)\citenamefont {Pastor},
  \citenamefont {Pinto},\ and\ \citenamefont {Raffelt}}]{pastor2009}%
  \BibitemOpen
  \bibfield  {author} {\bibinfo {author} {\bibfnamefont {S.}~\bibnamefont
  {Pastor}}, \bibinfo {author} {\bibfnamefont {T.}~\bibnamefont {Pinto}}, \
  and\ \bibinfo {author} {\bibfnamefont {G.~G.}\ \bibnamefont {Raffelt}},\
  }\href@noop {} {\bibfield  {journal} {\bibinfo  {journal} {Phys. Rev. Lett.}\
  }\textbf {\bibinfo {volume} {102}},\ \bibinfo {pages} {241302} (\bibinfo
  {year} {2009})}\BibitemShut {NoStop}%
\bibitem [{\citenamefont {Gava}\ and\ \citenamefont {Volpe}(2010)}]{gava2010}%
  \BibitemOpen
  \bibfield  {author} {\bibinfo {author} {\bibfnamefont {J.}~\bibnamefont
  {Gava}}\ and\ \bibinfo {author} {\bibfnamefont {C.}~\bibnamefont {Volpe}},\
  }\href@noop {} {\bibfield  {journal} {\bibinfo  {journal} {Nucl. Phys.}\
  }\textbf {\bibinfo {volume} {B837}},\ \bibinfo {pages} {50} (\bibinfo {year}
  {2010})}\BibitemShut {NoStop}%
\bibitem [{\citenamefont {Fuller}\ \emph {et~al.}(1987)\citenamefont {Fuller},
  \citenamefont {Mayle}, \citenamefont {Wilson},\ and\ \citenamefont
  {Schramm}}]{fuller1987}%
  \BibitemOpen
  \bibfield  {author} {\bibinfo {author} {\bibfnamefont {G.}~\bibnamefont
  {Fuller}}, \bibinfo {author} {\bibfnamefont {R.}~\bibnamefont {Mayle}},
  \bibinfo {author} {\bibfnamefont {J.}~\bibnamefont {Wilson}}, \ and\ \bibinfo
  {author} {\bibfnamefont {D.}~\bibnamefont {Schramm}},\ }\href@noop {}
  {\bibfield  {journal} {\bibinfo  {journal} {Astrophys. J.}\ }\textbf
  {\bibinfo {volume} {322}},\ \bibinfo {pages} {795} (\bibinfo {year}
  {1987})}\BibitemShut {NoStop}%
\bibitem [{\citenamefont {Fuller}\ \emph {et~al.}(1992)\citenamefont {Fuller},
  \citenamefont {Mayle}, \citenamefont {Meyer},\ and\ \citenamefont
  {Wilson}}]{fuller1992}%
  \BibitemOpen
  \bibfield  {author} {\bibinfo {author} {\bibfnamefont {G.~M.}\ \bibnamefont
  {Fuller}}, \bibinfo {author} {\bibfnamefont {R.}~\bibnamefont {Mayle}},
  \bibinfo {author} {\bibfnamefont {B.~S.}\ \bibnamefont {Meyer}}, \ and\
  \bibinfo {author} {\bibfnamefont {J.~R.}\ \bibnamefont {Wilson}},\
  }\href@noop {} {\bibfield  {journal} {\bibinfo  {journal} {Astrophys. J.}\
  }\textbf {\bibinfo {volume} {389}},\ \bibinfo {pages} {517} (\bibinfo {year}
  {1992})}\BibitemShut {NoStop}%
\bibitem [{\citenamefont {Qian}\ and\ \citenamefont
  {Fuller}(1995{\natexlab{a}})}]{qian1995a}%
  \BibitemOpen
  \bibfield  {author} {\bibinfo {author} {\bibfnamefont {Y.-Z.}\ \bibnamefont
  {Qian}}\ and\ \bibinfo {author} {\bibfnamefont {G.~M.}\ \bibnamefont
  {Fuller}},\ }\href@noop {} {\bibfield  {journal} {\bibinfo  {journal} {Phys.
  Rev. D}\ }\textbf {\bibinfo {volume} {51}},\ \bibinfo {pages} {1479}
  (\bibinfo {year} {1995}{\natexlab{a}})}\BibitemShut {NoStop}%
\bibitem [{\citenamefont {Qian}\ and\ \citenamefont
  {Fuller}(1995{\natexlab{b}})}]{qian1995b}%
  \BibitemOpen
  \bibfield  {author} {\bibinfo {author} {\bibfnamefont {Y.-Z.}\ \bibnamefont
  {Qian}}\ and\ \bibinfo {author} {\bibfnamefont {G.~M.}\ \bibnamefont
  {Fuller}},\ }\href@noop {} {\bibfield  {journal} {\bibinfo  {journal} {Phys.
  Rev. D}\ }\textbf {\bibinfo {volume} {52}},\ \bibinfo {pages} {656} (\bibinfo
  {year} {1995}{\natexlab{b}})}\BibitemShut {NoStop}%
\bibitem [{\citenamefont {Qian}\ \emph {et~al.}(1997)\citenamefont {Qian},
  \citenamefont {Haxton}, \citenamefont {Langanke},\ and\ \citenamefont
  {Vogel}}]{qian1997}%
  \BibitemOpen
  \bibfield  {author} {\bibinfo {author} {\bibfnamefont {Y.-Z.}\ \bibnamefont
  {Qian}}, \bibinfo {author} {\bibfnamefont {W.~C.}\ \bibnamefont {Haxton}},
  \bibinfo {author} {\bibfnamefont {K.}~\bibnamefont {Langanke}}, \ and\
  \bibinfo {author} {\bibfnamefont {P.}~\bibnamefont {Vogel}},\ }\href
  {\doibase 10.1103/PhysRevC.55.1532} {\bibfield  {journal} {\bibinfo
  {journal} {Phys. Rev. C}\ }\textbf {\bibinfo {volume} {55}},\ \bibinfo
  {pages} {1532} (\bibinfo {year} {1997})}\BibitemShut {NoStop}%
\bibitem [{\citenamefont {Horowitz}\ and\ \citenamefont
  {Li}(1999)}]{horowitz1999}%
  \BibitemOpen
  \bibfield  {author} {\bibinfo {author} {\bibfnamefont {C.~J.}\ \bibnamefont
  {Horowitz}}\ and\ \bibinfo {author} {\bibfnamefont {G.}~\bibnamefont {Li}},\
  }\href {\doibase 10.1103/PhysRevLett.82.5198} {\bibfield  {journal} {\bibinfo
   {journal} {Phys. Rev. Lett.}\ }\textbf {\bibinfo {volume} {82}},\ \bibinfo
  {pages} {5198} (\bibinfo {year} {1999})}\BibitemShut {NoStop}%
\bibitem [{\citenamefont {McLaughlin}\ \emph {et~al.}(1999)\citenamefont
  {McLaughlin}, \citenamefont {Fetter}, \citenamefont {Balantekin},\ and\
  \citenamefont {Fuller}}]{mclaughlin1999}%
  \BibitemOpen
  \bibfield  {author} {\bibinfo {author} {\bibfnamefont {G.~C.}\ \bibnamefont
  {McLaughlin}}, \bibinfo {author} {\bibfnamefont {J.~M.}\ \bibnamefont
  {Fetter}}, \bibinfo {author} {\bibfnamefont {A.~B.}\ \bibnamefont
  {Balantekin}}, \ and\ \bibinfo {author} {\bibfnamefont {G.~M.}\ \bibnamefont
  {Fuller}},\ }\href {\doibase 10.1103/PhysRevC.59.2873} {\bibfield  {journal}
  {\bibinfo  {journal} {Phys. Rev. C}\ }\textbf {\bibinfo {volume} {59}},\
  \bibinfo {pages} {2873} (\bibinfo {year} {1999})}\BibitemShut {NoStop}%
\bibitem [{\citenamefont {Schirato}\ and\ \citenamefont
  {Fuller}(2002)}]{schirato2002}%
  \BibitemOpen
  \bibfield  {author} {\bibinfo {author} {\bibfnamefont {R.~C.}\ \bibnamefont
  {Schirato}}\ and\ \bibinfo {author} {\bibfnamefont {G.~M.}\ \bibnamefont
  {Fuller}},\ }\href@noop {} {\bibfield  {journal} {\bibinfo  {journal}
  {arXiv:0205390}\ } (\bibinfo {year} {2002})}\BibitemShut {NoStop}%
\bibitem [{\citenamefont {Pastor}\ \emph {et~al.}(2002)\citenamefont {Pastor},
  \citenamefont {Raffelt},\ and\ \citenamefont {Semikoz}}]{pastor2002}%
  \BibitemOpen
  \bibfield  {author} {\bibinfo {author} {\bibfnamefont {S.}~\bibnamefont
  {Pastor}}, \bibinfo {author} {\bibfnamefont {G.}~\bibnamefont {Raffelt}}, \
  and\ \bibinfo {author} {\bibfnamefont {D.~V.}\ \bibnamefont {Semikoz}},\
  }\href {\doibase 10.1103/PhysRevD.65.053011} {\bibfield  {journal} {\bibinfo
  {journal} {Phys. Rev. D}\ }\textbf {\bibinfo {volume} {65}},\ \bibinfo
  {pages} {053011} (\bibinfo {year} {2002})}\BibitemShut {NoStop}%
\bibitem [{\citenamefont {{Pastor}}\ and\ \citenamefont
  {{Raffelt}}(2002)}]{pastor2002b}%
  \BibitemOpen
  \bibfield  {author} {\bibinfo {author} {\bibfnamefont {S.}~\bibnamefont
  {{Pastor}}}\ and\ \bibinfo {author} {\bibfnamefont {G.}~\bibnamefont
  {{Raffelt}}},\ }\href {\doibase 10.1103/PhysRevLett.89.191101} {\bibfield
  {journal} {\bibinfo  {journal} {Phys. Rev. Lett.}\ }\textbf {\bibinfo
  {volume} {89}},\ \bibinfo {pages} {191101} (\bibinfo {year}
  {2002})}\BibitemShut {NoStop}%
\bibitem [{\citenamefont {Fetter}\ \emph {et~al.}(2003)\citenamefont {Fetter},
  \citenamefont {McLaughlin}, \citenamefont {Balantekin},\ and\ \citenamefont
  {Fuller}}]{fetter2003}%
  \BibitemOpen
  \bibfield  {author} {\bibinfo {author} {\bibfnamefont {J.}~\bibnamefont
  {Fetter}}, \bibinfo {author} {\bibfnamefont {G.~C.}\ \bibnamefont
  {McLaughlin}}, \bibinfo {author} {\bibfnamefont {A.~B.}\ \bibnamefont
  {Balantekin}}, \ and\ \bibinfo {author} {\bibfnamefont {G.~M.}\ \bibnamefont
  {Fuller}},\ }\href {\doibase https://doi.org/10.1016/S0927-6505(02)00156-1}
  {\bibfield  {journal} {\bibinfo  {journal} {Astropart. Phys.}\ }\textbf
  {\bibinfo {volume} {18}},\ \bibinfo {pages} {433 } (\bibinfo {year}
  {2003})}\BibitemShut {NoStop}%
\bibitem [{\citenamefont {{Duan}}\ \emph {et~al.}(2006)\citenamefont {{Duan}},
  \citenamefont {{Fuller}}, \citenamefont {{Carlson}},\ and\ \citenamefont
  {{Qian}}}]{duan2006b}%
  \BibitemOpen
  \bibfield  {author} {\bibinfo {author} {\bibfnamefont {H.}~\bibnamefont
  {{Duan}}}, \bibinfo {author} {\bibfnamefont {G.~M.}\ \bibnamefont
  {{Fuller}}}, \bibinfo {author} {\bibfnamefont {J.}~\bibnamefont {{Carlson}}},
  \ and\ \bibinfo {author} {\bibfnamefont {Y.-Z.}\ \bibnamefont {{Qian}}},\
  }\href {\doibase 10.1103/PhysRevLett.97.241101} {\bibfield  {journal}
  {\bibinfo  {journal} {Phys. Rev. Lett.}\ }\textbf {\bibinfo {volume} {97}},\
  \bibinfo {pages} {241101} (\bibinfo {year} {2006})}\BibitemShut {NoStop}%
\bibitem [{\citenamefont {Duan}\ \emph {et~al.}(2011)\citenamefont {Duan},
  \citenamefont {Friedland}, \citenamefont {McLaughlin},\ and\ \citenamefont
  {Surman}}]{duan2011}%
  \BibitemOpen
  \bibfield  {author} {\bibinfo {author} {\bibfnamefont {H.}~\bibnamefont
  {Duan}}, \bibinfo {author} {\bibfnamefont {A.}~\bibnamefont {Friedland}},
  \bibinfo {author} {\bibfnamefont {G.~C.}\ \bibnamefont {McLaughlin}}, \ and\
  \bibinfo {author} {\bibfnamefont {R.}~\bibnamefont {Surman}},\ }\href@noop {}
  {\bibfield  {journal} {\bibinfo  {journal} {J. Phys. G}\ }\textbf {\bibinfo
  {volume} {38}},\ \bibinfo {pages} {035201} (\bibinfo {year}
  {2011})}\BibitemShut {NoStop}%
\bibitem [{\citenamefont {Tamborra}\ \emph {et~al.}(2012)\citenamefont
  {Tamborra}, \citenamefont {Raffelt}, \citenamefont {H{\"u}depohl},\ and\
  \citenamefont {Janka}}]{tamborra2012}%
  \BibitemOpen
  \bibfield  {author} {\bibinfo {author} {\bibfnamefont {I.}~\bibnamefont
  {Tamborra}}, \bibinfo {author} {\bibfnamefont {G.~G.}\ \bibnamefont
  {Raffelt}}, \bibinfo {author} {\bibfnamefont {L.}~\bibnamefont
  {H{\"u}depohl}}, \ and\ \bibinfo {author} {\bibfnamefont {H.-T.}\
  \bibnamefont {Janka}},\ }\href
  {http://stacks.iop.org/1475-7516/2012/i=01/a=013} {\bibfield  {journal}
  {\bibinfo  {journal} {J. Cosmol. Astropart. Phys.}\ }\textbf {\bibinfo
  {volume} {2012}},\ \bibinfo {pages} {013} (\bibinfo {year}
  {2012})}\BibitemShut {NoStop}%
\bibitem [{\citenamefont {Malkus}\ \emph {et~al.}(2012)\citenamefont {Malkus},
  \citenamefont {Kneller}, \citenamefont {McLaughlin},\ and\ \citenamefont
  {Surman}}]{malkus2012}%
  \BibitemOpen
  \bibfield  {author} {\bibinfo {author} {\bibfnamefont {A.}~\bibnamefont
  {Malkus}}, \bibinfo {author} {\bibfnamefont {J.~P.}\ \bibnamefont {Kneller}},
  \bibinfo {author} {\bibfnamefont {G.~C.}\ \bibnamefont {McLaughlin}}, \ and\
  \bibinfo {author} {\bibfnamefont {R.}~\bibnamefont {Surman}},\ }\href@noop {}
  {\bibfield  {journal} {\bibinfo  {journal} {Phys. Rev. D}\ }\textbf {\bibinfo
  {volume} {86}},\ \bibinfo {pages} {085015} (\bibinfo {year}
  {2012})}\BibitemShut {NoStop}%
\bibitem [{\citenamefont {Wu}\ \emph {et~al.}(2015)\citenamefont {Wu},
  \citenamefont {Qian}, \citenamefont {Mart\'{\i}nez-Pinedo}, \citenamefont
  {Fischer},\ and\ \citenamefont {Huther}}]{wu2015}%
  \BibitemOpen
  \bibfield  {author} {\bibinfo {author} {\bibfnamefont {M.-R.}\ \bibnamefont
  {Wu}}, \bibinfo {author} {\bibfnamefont {Y.-Z.}\ \bibnamefont {Qian}},
  \bibinfo {author} {\bibfnamefont {G.}~\bibnamefont {Mart\'{\i}nez-Pinedo}},
  \bibinfo {author} {\bibfnamefont {T.}~\bibnamefont {Fischer}}, \ and\
  \bibinfo {author} {\bibfnamefont {L.}~\bibnamefont {Huther}},\ }\href
  {\doibase 10.1103/PhysRevD.91.065016} {\bibfield  {journal} {\bibinfo
  {journal} {Phys. Rev. D}\ }\textbf {\bibinfo {volume} {91}},\ \bibinfo
  {pages} {065016} (\bibinfo {year} {2015})}\BibitemShut {NoStop}%
\bibitem [{\citenamefont {Dighe}\ and\ \citenamefont
  {Smirnov}(2000)}]{dighe2000}%
  \BibitemOpen
  \bibfield  {author} {\bibinfo {author} {\bibfnamefont {A.~S.}\ \bibnamefont
  {Dighe}}\ and\ \bibinfo {author} {\bibfnamefont {A.~Y.}\ \bibnamefont
  {Smirnov}},\ }\href {\doibase 10.1103/PhysRevD.62.033007} {\bibfield
  {journal} {\bibinfo  {journal} {Phys. Rev. D}\ }\textbf {\bibinfo {volume}
  {62}},\ \bibinfo {pages} {033007} (\bibinfo {year} {2000})}\BibitemShut
  {NoStop}%
\bibitem [{\citenamefont {Abazajian}\ \emph {et~al.}(2011)\citenamefont
  {Abazajian}, \citenamefont {Calabrese}, \citenamefont {Cooray}, \citenamefont
  {Bernardis}, \citenamefont {Dodelson} \emph {et~al.}}]{abazajian2011}%
  \BibitemOpen
  \bibfield  {author} {\bibinfo {author} {\bibfnamefont {K.}~\bibnamefont
  {Abazajian}}, \bibinfo {author} {\bibfnamefont {E.}~\bibnamefont
  {Calabrese}}, \bibinfo {author} {\bibfnamefont {A.}~\bibnamefont {Cooray}},
  \bibinfo {author} {\bibfnamefont {F.~D.}\ \bibnamefont {Bernardis}}, \bibinfo
  {author} {\bibfnamefont {S.}~\bibnamefont {Dodelson}},  \emph {et~al.},\
  }\href {\doibase http://dx.doi.org/10.1016/j.astropartphys.2011.07.002}
  {\bibfield  {journal} {\bibinfo  {journal} {Astropart. Phys.}\ }\textbf
  {\bibinfo {volume} {35}},\ \bibinfo {pages} {177 } (\bibinfo {year}
  {2011})}\BibitemShut {NoStop}%
\bibitem [{\citenamefont {Scholberg}(2017)}]{scholberg2017}%
  \BibitemOpen
  \bibfield  {author} {\bibinfo {author} {\bibfnamefont {K.}~\bibnamefont
  {Scholberg}},\ }\href@noop {} {\bibfield  {journal} {\bibinfo  {journal}
  {arXiv:1707.06384}\ } (\bibinfo {year} {2017})}\BibitemShut {NoStop}%
\bibitem [{\citenamefont {Sawyer}(2005)}]{sawyer2005}%
  \BibitemOpen
  \bibfield  {author} {\bibinfo {author} {\bibfnamefont {R.~F.}\ \bibnamefont
  {Sawyer}},\ }\href {\doibase 10.1103/PhysRevD.72.045003} {\bibfield
  {journal} {\bibinfo  {journal} {Phys. Rev. D}\ }\textbf {\bibinfo {volume}
  {72}},\ \bibinfo {pages} {045003} (\bibinfo {year} {2005})}\BibitemShut
  {NoStop}%
\bibitem [{\citenamefont {Cherry}\ \emph {et~al.}(2012)\citenamefont {Cherry},
  \citenamefont {Carlson}, \citenamefont {Friedland}, \citenamefont {Fuller},\
  and\ \citenamefont {Vlasenko}}]{cherry2012}%
  \BibitemOpen
  \bibfield  {author} {\bibinfo {author} {\bibfnamefont {J.~F.}\ \bibnamefont
  {Cherry}}, \bibinfo {author} {\bibfnamefont {J.}~\bibnamefont {Carlson}},
  \bibinfo {author} {\bibfnamefont {A.}~\bibnamefont {Friedland}}, \bibinfo
  {author} {\bibfnamefont {G.~M.}\ \bibnamefont {Fuller}}, \ and\ \bibinfo
  {author} {\bibfnamefont {A.}~\bibnamefont {Vlasenko}},\ }\href@noop {}
  {\bibfield  {journal} {\bibinfo  {journal} {Phys. Rev. Lett.}\ }\textbf
  {\bibinfo {volume} {108}},\ \bibinfo {pages} {261104} (\bibinfo {year}
  {2012})}\BibitemShut {NoStop}%
\bibitem [{\citenamefont {Raffelt}\ \emph {et~al.}(2013)\citenamefont
  {Raffelt}, \citenamefont {Sarikas},\ and\ \citenamefont
  {Seixas}}]{raffelt2013}%
  \BibitemOpen
  \bibfield  {author} {\bibinfo {author} {\bibfnamefont {G.}~\bibnamefont
  {Raffelt}}, \bibinfo {author} {\bibfnamefont {S.}~\bibnamefont {Sarikas}}, \
  and\ \bibinfo {author} {\bibfnamefont {D.~S.}\ \bibnamefont {Seixas}},\
  }\href {\doibase 10.1103/PhysRevLett.111.091101} {\bibfield  {journal}
  {\bibinfo  {journal} {Phys. Rev. Lett.}\ }\textbf {\bibinfo {volume} {111}},\
  \bibinfo {pages} {091101} (\bibinfo {year} {2013})}\BibitemShut {NoStop}%
\bibitem [{\citenamefont {Hansen}\ and\ \citenamefont
  {Hannestad}(2014)}]{hansen2014}%
  \BibitemOpen
  \bibfield  {author} {\bibinfo {author} {\bibfnamefont {R.~S.}\ \bibnamefont
  {Hansen}}\ and\ \bibinfo {author} {\bibfnamefont {S.}~\bibnamefont
  {Hannestad}},\ }\href {\doibase 10.1103/PhysRevD.90.025009} {\bibfield
  {journal} {\bibinfo  {journal} {Phys. Rev. D}\ }\textbf {\bibinfo {volume}
  {90}},\ \bibinfo {pages} {025009} (\bibinfo {year} {2014})}\BibitemShut
  {NoStop}%
\bibitem [{\citenamefont {Vlasenko}\ \emph {et~al.}()\citenamefont {Vlasenko},
  \citenamefont {Fuller},\ and\ \citenamefont {Cirigliano}}]{vlasenko2014b}%
  \BibitemOpen
  \bibfield  {author} {\bibinfo {author} {\bibfnamefont {A.}~\bibnamefont
  {Vlasenko}}, \bibinfo {author} {\bibfnamefont {G.~M.}\ \bibnamefont
  {Fuller}}, \ and\ \bibinfo {author} {\bibfnamefont {V.}~\bibnamefont
  {Cirigliano}},\ }\href@noop {} {\bibinfo  {journal} {arXiv:1406.6724}\
  }\BibitemShut {NoStop}%
\bibitem [{\citenamefont {Serreau}\ and\ \citenamefont
  {Volpe}(2014)}]{serreau2014}%
  \BibitemOpen
\bibfield  {journal} {  }\bibfield  {author} {\bibinfo {author} {\bibfnamefont
  {J.}~\bibnamefont {Serreau}}\ and\ \bibinfo {author} {\bibfnamefont
  {C.}~\bibnamefont {Volpe}},\ }\href {\doibase 10.1103/PhysRevD.90.125040}
  {\bibfield  {journal} {\bibinfo  {journal} {Phys. Rev. D}\ }\textbf {\bibinfo
  {volume} {90}},\ \bibinfo {pages} {125040} (\bibinfo {year}
  {2014})}\BibitemShut {NoStop}%
\bibitem [{\citenamefont {Abbar}\ \emph {et~al.}(2015)\citenamefont {Abbar},
  \citenamefont {Duan},\ and\ \citenamefont {Shalgar}}]{abbar2015}%
  \BibitemOpen
  \bibfield  {author} {\bibinfo {author} {\bibfnamefont {S.}~\bibnamefont
  {Abbar}}, \bibinfo {author} {\bibfnamefont {H.}~\bibnamefont {Duan}}, \ and\
  \bibinfo {author} {\bibfnamefont {S.}~\bibnamefont {Shalgar}},\ }\href
  {\doibase 10.1103/PhysRevD.92.065019} {\bibfield  {journal} {\bibinfo
  {journal} {Phys. Rev. D}\ }\textbf {\bibinfo {volume} {92}},\ \bibinfo
  {pages} {065019} (\bibinfo {year} {2015})}\BibitemShut {NoStop}%
\bibitem [{\citenamefont {Mirizzi}\ \emph {et~al.}(2015)\citenamefont
  {Mirizzi}, \citenamefont {Mangano},\ and\ \citenamefont
  {Saviano}}]{mirizzi2015}%
  \BibitemOpen
  \bibfield  {author} {\bibinfo {author} {\bibfnamefont {A.}~\bibnamefont
  {Mirizzi}}, \bibinfo {author} {\bibfnamefont {G.}~\bibnamefont {Mangano}}, \
  and\ \bibinfo {author} {\bibfnamefont {N.}~\bibnamefont {Saviano}},\ }\href
  {\doibase 10.1103/PhysRevD.92.021702} {\bibfield  {journal} {\bibinfo
  {journal} {Phys. Rev. D}\ }\textbf {\bibinfo {volume} {92}},\ \bibinfo
  {pages} {021702} (\bibinfo {year} {2015})}\BibitemShut {NoStop}%
\bibitem [{\citenamefont {Keister}(2015)}]{keister2015}%
  \BibitemOpen
  \bibfield  {author} {\bibinfo {author} {\bibfnamefont {B.~D.}\ \bibnamefont
  {Keister}},\ }\href {http://stacks.iop.org/1402-4896/90/i=8/a=088008}
  {\bibfield  {journal} {\bibinfo  {journal} {Phys. Scripta}\ }\textbf
  {\bibinfo {volume} {90}},\ \bibinfo {pages} {088008} (\bibinfo {year}
  {2015})}\BibitemShut {NoStop}%
\bibitem [{\citenamefont {Armstrong}\ \emph {et~al.}(2017)\citenamefont
  {Armstrong}, \citenamefont {Patwardhan}, \citenamefont {Johns}, \citenamefont
  {Kishimoto}, \citenamefont {Abarbanel},\ and\ \citenamefont
  {Fuller}}]{armstrong2016}%
  \BibitemOpen
  \bibfield  {author} {\bibinfo {author} {\bibfnamefont {E.}~\bibnamefont
  {Armstrong}}, \bibinfo {author} {\bibfnamefont {A.~V.}\ \bibnamefont
  {Patwardhan}}, \bibinfo {author} {\bibfnamefont {L.}~\bibnamefont {Johns}},
  \bibinfo {author} {\bibfnamefont {C.~T.}\ \bibnamefont {Kishimoto}}, \bibinfo
  {author} {\bibfnamefont {H.~D.~I.}\ \bibnamefont {Abarbanel}}, \ and\
  \bibinfo {author} {\bibfnamefont {G.~M.}\ \bibnamefont {Fuller}},\ }\href
  {\doibase 10.1103/PhysRevD.96.083008} {\bibfield  {journal} {\bibinfo
  {journal} {Phys. Rev. D}\ }\textbf {\bibinfo {volume} {96}},\ \bibinfo
  {pages} {083008} (\bibinfo {year} {2017})}\BibitemShut {NoStop}%
\bibitem [{\citenamefont {Chakraborty}\ \emph {et~al.}(2016)\citenamefont
  {Chakraborty}, \citenamefont {Hansen}, \citenamefont {Izaguirre},\ and\
  \citenamefont {Raffelt}}]{chakraborty2016}%
  \BibitemOpen
  \bibfield  {author} {\bibinfo {author} {\bibfnamefont {S.}~\bibnamefont
  {Chakraborty}}, \bibinfo {author} {\bibfnamefont {R.}~\bibnamefont {Hansen}},
  \bibinfo {author} {\bibfnamefont {I.}~\bibnamefont {Izaguirre}}, \ and\
  \bibinfo {author} {\bibfnamefont {G.}~\bibnamefont {Raffelt}},\ }\href
  {\doibase http://dx.doi.org/10.1016/j.nuclphysb.2016.02.012} {\bibfield
  {journal} {\bibinfo  {journal} {Nucl. Phys.}\ }\textbf {\bibinfo {volume}
  {B908}},\ \bibinfo {pages} {366 } (\bibinfo {year} {2016})}\BibitemShut
  {NoStop}%
\bibitem [{\citenamefont {Sawyer}(2016)}]{sawyer2016}%
  \BibitemOpen
  \bibfield  {author} {\bibinfo {author} {\bibfnamefont {R.~F.}\ \bibnamefont
  {Sawyer}},\ }\href {\doibase 10.1103/PhysRevLett.116.081101} {\bibfield
  {journal} {\bibinfo  {journal} {Phys. Rev. Lett.}\ }\textbf {\bibinfo
  {volume} {116}},\ \bibinfo {pages} {081101} (\bibinfo {year}
  {2016})}\BibitemShut {NoStop}%
\bibitem [{\citenamefont {Volpe}(2016)}]{volpe2016}%
  \BibitemOpen
  \bibfield  {author} {\bibinfo {author} {\bibfnamefont {C.}~\bibnamefont
  {Volpe}},\ }\href {http://stacks.iop.org/1742-6596/718/i=6/a=062068}
  {\bibfield  {journal} {\bibinfo  {journal} {J. Phys. Conf. Ser.}\ }\textbf
  {\bibinfo {volume} {718}},\ \bibinfo {pages} {062068} (\bibinfo {year}
  {2016})}\BibitemShut {NoStop}%
\bibitem [{\citenamefont {Johns}\ and\ \citenamefont
  {Fuller}(2017)}]{johns2017}%
  \BibitemOpen
  \bibfield  {author} {\bibinfo {author} {\bibfnamefont {L.}~\bibnamefont
  {Johns}}\ and\ \bibinfo {author} {\bibfnamefont {G.~M.}\ \bibnamefont
  {Fuller}},\ }\href {\doibase 10.1103/PhysRevD.95.043003} {\bibfield
  {journal} {\bibinfo  {journal} {Phys. Rev. D}\ }\textbf {\bibinfo {volume}
  {95}},\ \bibinfo {pages} {043003} (\bibinfo {year} {2017})}\BibitemShut
  {NoStop}%
\bibitem [{\citenamefont {Dasgupta}\ \emph {et~al.}(2017)\citenamefont
  {Dasgupta}, \citenamefont {Mirizzi},\ and\ \citenamefont
  {Sen}}]{dasgupta2017}%
  \BibitemOpen
  \bibfield  {author} {\bibinfo {author} {\bibfnamefont {B.}~\bibnamefont
  {Dasgupta}}, \bibinfo {author} {\bibfnamefont {A.}~\bibnamefont {Mirizzi}}, \
  and\ \bibinfo {author} {\bibfnamefont {M.}~\bibnamefont {Sen}},\ }\href
  {http://stacks.iop.org/1475-7516/2017/i=02/a=019} {\bibfield  {journal}
  {\bibinfo  {journal} {J. Cosmol. Astropart. Phys.}\ }\textbf {\bibinfo
  {volume} {2017}},\ \bibinfo {pages} {019} (\bibinfo {year}
  {2017})}\BibitemShut {NoStop}%
\bibitem [{\citenamefont {Tian}\ \emph {et~al.}(2017)\citenamefont {Tian},
  \citenamefont {Patwardhan},\ and\ \citenamefont {Fuller}}]{tian2017}%
  \BibitemOpen
  \bibfield  {author} {\bibinfo {author} {\bibfnamefont {J.~Y.}\ \bibnamefont
  {Tian}}, \bibinfo {author} {\bibfnamefont {A.~V.}\ \bibnamefont
  {Patwardhan}}, \ and\ \bibinfo {author} {\bibfnamefont {G.~M.}\ \bibnamefont
  {Fuller}},\ }\href {\doibase 10.1103/PhysRevD.95.063004} {\bibfield
  {journal} {\bibinfo  {journal} {Phys. Rev. D}\ }\textbf {\bibinfo {volume}
  {95}},\ \bibinfo {pages} {063004} (\bibinfo {year} {2017})}\BibitemShut
  {NoStop}%
\bibitem [{\citenamefont {Wu}\ and\ \citenamefont {Tamborra}(2017)}]{wu2017}%
  \BibitemOpen
  \bibfield  {author} {\bibinfo {author} {\bibfnamefont {M.-R.}\ \bibnamefont
  {Wu}}\ and\ \bibinfo {author} {\bibfnamefont {I.}~\bibnamefont {Tamborra}},\
  }\href {\doibase 10.1103/PhysRevD.95.103007} {\bibfield  {journal} {\bibinfo
  {journal} {Phys. Rev. D}\ }\textbf {\bibinfo {volume} {95}},\ \bibinfo
  {pages} {103007} (\bibinfo {year} {2017})}\BibitemShut {NoStop}%
\bibitem [{\citenamefont {Izaguirre}\ \emph {et~al.}(2017)\citenamefont
  {Izaguirre}, \citenamefont {Raffelt},\ and\ \citenamefont
  {Tamborra}}]{izaguirre2017}%
  \BibitemOpen
  \bibfield  {author} {\bibinfo {author} {\bibfnamefont {I.}~\bibnamefont
  {Izaguirre}}, \bibinfo {author} {\bibfnamefont {G.}~\bibnamefont {Raffelt}},
  \ and\ \bibinfo {author} {\bibfnamefont {I.}~\bibnamefont {Tamborra}},\
  }\href {\doibase 10.1103/PhysRevLett.118.021101} {\bibfield  {journal}
  {\bibinfo  {journal} {Phys. Rev. Lett.}\ }\textbf {\bibinfo {volume} {118}},\
  \bibinfo {pages} {021101} (\bibinfo {year} {2017})}\BibitemShut {NoStop}%
\bibitem [{\citenamefont {Cirigliano}\ \emph {et~al.}(2017)\citenamefont
  {Cirigliano}, \citenamefont {Paris},\ and\ \citenamefont
  {Shalgar}}]{cirigliano2017}%
  \BibitemOpen
  \bibfield  {author} {\bibinfo {author} {\bibfnamefont {V.}~\bibnamefont
  {Cirigliano}}, \bibinfo {author} {\bibfnamefont {M.~W.}\ \bibnamefont
  {Paris}}, \ and\ \bibinfo {author} {\bibfnamefont {S.}~\bibnamefont
  {Shalgar}},\ }\href {\doibase https://doi.org/10.1016/j.physletb.2017.09.039}
  {\bibfield  {journal} {\bibinfo  {journal} {Phys. Lett. B}\ }\textbf
  {\bibinfo {volume} {774}},\ \bibinfo {pages} {258 } (\bibinfo {year}
  {2017})}\BibitemShut {NoStop}%
\bibitem [{\citenamefont {{Hannestad}}\ \emph {et~al.}(2006)\citenamefont
  {{Hannestad}}, \citenamefont {{Raffelt}}, \citenamefont {{Sigl}},\ and\
  \citenamefont {{Wong}}}]{hannestad2006}%
  \BibitemOpen
  \bibfield  {author} {\bibinfo {author} {\bibfnamefont {S.}~\bibnamefont
  {{Hannestad}}}, \bibinfo {author} {\bibfnamefont {G.~G.}\ \bibnamefont
  {{Raffelt}}}, \bibinfo {author} {\bibfnamefont {G.}~\bibnamefont {{Sigl}}}, \
  and\ \bibinfo {author} {\bibfnamefont {Y.~Y.~Y.}\ \bibnamefont {{Wong}}},\
  }\href {\doibase 10.1103/PhysRevD.74.105010} {\bibfield  {journal} {\bibinfo
  {journal} {Phys. Rev. D}\ }\textbf {\bibinfo {volume} {74}},\ \bibinfo
  {pages} {105010} (\bibinfo {year} {2006})}\BibitemShut {NoStop}%
\bibitem [{\citenamefont {{Duan}}\ \emph {et~al.}(2007)\citenamefont {{Duan}},
  \citenamefont {{Fuller}}, \citenamefont {{Carlson}},\ and\ \citenamefont
  {{Qian}}}]{duan2007b}%
  \BibitemOpen
  \bibfield  {author} {\bibinfo {author} {\bibfnamefont {H.}~\bibnamefont
  {{Duan}}}, \bibinfo {author} {\bibfnamefont {G.~M.}\ \bibnamefont
  {{Fuller}}}, \bibinfo {author} {\bibfnamefont {J.}~\bibnamefont {{Carlson}}},
  \ and\ \bibinfo {author} {\bibfnamefont {Y.-Z.}\ \bibnamefont {{Qian}}},\
  }\href {\doibase 10.1103/PhysRevD.75.125005} {\bibfield  {journal} {\bibinfo
  {journal} {Phys. Rev. D}\ }\textbf {\bibinfo {volume} {75}},\ \bibinfo
  {pages} {125005} (\bibinfo {year} {2007})}\BibitemShut {NoStop}%
\bibitem [{\citenamefont {Wolfenstein}(1978)}]{wolfenstein1978}%
  \BibitemOpen
  \bibfield  {author} {\bibinfo {author} {\bibfnamefont {L.}~\bibnamefont
  {Wolfenstein}},\ }\href@noop {} {\bibfield  {journal} {\bibinfo  {journal}
  {Phys. Rev. D}\ }\textbf {\bibinfo {volume} {17}},\ \bibinfo {pages} {2369}
  (\bibinfo {year} {1978})}\BibitemShut {NoStop}%
\bibitem [{\citenamefont {Mikheyev}\ and\ \citenamefont
  {Smirnov}(1985)}]{mikheyev1985}%
  \BibitemOpen
  \bibfield  {author} {\bibinfo {author} {\bibfnamefont {S.~P.}\ \bibnamefont
  {Mikheyev}}\ and\ \bibinfo {author} {\bibfnamefont {A.~Y.}\ \bibnamefont
  {Smirnov}},\ }\href@noop {} {\bibfield  {journal} {\bibinfo  {journal} {Sov.
  J. Nucl. Phys.}\ }\textbf {\bibinfo {volume} {42}},\ \bibinfo {pages} {913 }
  (\bibinfo {year} {1985})}\BibitemShut {NoStop}%
\bibitem [{\citenamefont {Duan}\ \emph {et~al.}(2010)\citenamefont {Duan},
  \citenamefont {Fuller},\ and\ \citenamefont {Qian}}]{duan2010}%
  \BibitemOpen
  \bibfield  {author} {\bibinfo {author} {\bibfnamefont {H.}~\bibnamefont
  {Duan}}, \bibinfo {author} {\bibfnamefont {G.~M.}\ \bibnamefont {Fuller}}, \
  and\ \bibinfo {author} {\bibfnamefont {Y.-Z.}\ \bibnamefont {Qian}},\
  }\href@noop {} {\bibfield  {journal} {\bibinfo  {journal} {Ann. Rev. Nucl.
  Part. Sci.}\ }\textbf {\bibinfo {volume} {60}},\ \bibinfo {pages} {569}
  (\bibinfo {year} {2010})}\BibitemShut {NoStop}%
\bibitem [{\citenamefont {Kosteleck\'y}\ and\ \citenamefont
  {Samuel}(1995)}]{kostelecky1995}%
  \BibitemOpen
  \bibfield  {author} {\bibinfo {author} {\bibfnamefont {V.~A.}\ \bibnamefont
  {Kosteleck\'y}}\ and\ \bibinfo {author} {\bibfnamefont {S.}~\bibnamefont
  {Samuel}},\ }\href {\doibase 10.1103/PhysRevD.52.621} {\bibfield  {journal}
  {\bibinfo  {journal} {Phys. Rev. D}\ }\textbf {\bibinfo {volume} {52}},\
  \bibinfo {pages} {621} (\bibinfo {year} {1995})}\BibitemShut {NoStop}%
\bibitem [{\citenamefont {Samuel}(1996)}]{samuel1996}%
  \BibitemOpen
  \bibfield  {author} {\bibinfo {author} {\bibfnamefont {S.}~\bibnamefont
  {Samuel}},\ }\href {\doibase 10.1103/PhysRevD.53.5382} {\bibfield  {journal}
  {\bibinfo  {journal} {Phys. Rev. D}\ }\textbf {\bibinfo {volume} {53}},\
  \bibinfo {pages} {5382} (\bibinfo {year} {1996})}\BibitemShut {NoStop}%
\bibitem [{\citenamefont {Steigman}(2012)}]{steigman2012}%
  \BibitemOpen
  \bibfield  {author} {\bibinfo {author} {\bibfnamefont {G.}~\bibnamefont
  {Steigman}},\ }\href@noop {} {\bibfield  {journal} {\bibinfo  {journal} {Adv.
  High Energy Phys.}\ }\textbf {\bibinfo {volume} {2012}},\ \bibinfo {pages}
  {268321} (\bibinfo {year} {2012})}\BibitemShut {NoStop}%
\bibitem [{\citenamefont {Castorina}\ \emph {et~al.}(2012)\citenamefont
  {Castorina}, \citenamefont {Franca}, \citenamefont {Lattanzi}, \citenamefont
  {Lesgourgues}, \citenamefont {Mangano}, \citenamefont {Melchiorri},\ and\
  \citenamefont {Pastor}}]{castorina2012}%
  \BibitemOpen
  \bibfield  {author} {\bibinfo {author} {\bibfnamefont {E.}~\bibnamefont
  {Castorina}}, \bibinfo {author} {\bibfnamefont {U.}~\bibnamefont {Franca}},
  \bibinfo {author} {\bibfnamefont {M.}~\bibnamefont {Lattanzi}}, \bibinfo
  {author} {\bibfnamefont {J.}~\bibnamefont {Lesgourgues}}, \bibinfo {author}
  {\bibfnamefont {G.}~\bibnamefont {Mangano}}, \bibinfo {author} {\bibfnamefont
  {A.}~\bibnamefont {Melchiorri}}, \ and\ \bibinfo {author} {\bibfnamefont
  {S.}~\bibnamefont {Pastor}},\ }\href {\doibase 10.1103/PhysRevD.86.023517}
  {\bibfield  {journal} {\bibinfo  {journal} {Phys. Rev. D}\ }\textbf {\bibinfo
  {volume} {86}},\ \bibinfo {pages} {023517} (\bibinfo {year}
  {2012})}\BibitemShut {NoStop}%
\bibitem [{\citenamefont {Barenboim}\ \emph {et~al.}(2017)\citenamefont
  {Barenboim}, \citenamefont {Kinney},\ and\ \citenamefont
  {Park}}]{barenboim2017}%
  \BibitemOpen
  \bibfield  {author} {\bibinfo {author} {\bibfnamefont {G.}~\bibnamefont
  {Barenboim}}, \bibinfo {author} {\bibfnamefont {W.~H.}\ \bibnamefont
  {Kinney}}, \ and\ \bibinfo {author} {\bibfnamefont {W.-I.}\ \bibnamefont
  {Park}},\ }\href {\doibase 10.1103/PhysRevD.95.043506} {\bibfield  {journal}
  {\bibinfo  {journal} {Phys. Rev. D}\ }\textbf {\bibinfo {volume} {95}},\
  \bibinfo {pages} {043506} (\bibinfo {year} {2017})}\BibitemShut {NoStop}%
\bibitem [{\citenamefont {Eijima}\ and\ \citenamefont
  {Shaposhnikov}(2017)}]{eijima2017}%
  \BibitemOpen
  \bibfield  {author} {\bibinfo {author} {\bibfnamefont {S.}~\bibnamefont
  {Eijima}}\ and\ \bibinfo {author} {\bibfnamefont {M.}~\bibnamefont
  {Shaposhnikov}},\ }\href {\doibase
  http://dx.doi.org/10.1016/j.physletb.2017.05.068} {\bibfield  {journal}
  {\bibinfo  {journal} {Phys. Lett. B}\ }\textbf {\bibinfo {volume} {771}},\
  \bibinfo {pages} {288 } (\bibinfo {year} {2017})}\BibitemShut {NoStop}%
\bibitem [{\citenamefont {Case}\ and\ \citenamefont {Jalal}(2014)}]{case2014}%
  \BibitemOpen
  \bibfield  {author} {\bibinfo {author} {\bibfnamefont {W.}~\bibnamefont
  {Case}}\ and\ \bibinfo {author} {\bibfnamefont {S.}~\bibnamefont {Jalal}},\
  }\href@noop {} {\bibfield  {journal} {\bibinfo  {journal} {Am. J. Phys.}\
  }\textbf {\bibinfo {volume} {82}},\ \bibinfo {pages} {654} (\bibinfo {year}
  {2014})}\BibitemShut {NoStop}%
\bibitem [{\citenamefont {Abazajian}(2014)}]{abazajian2014}%
  \BibitemOpen
  \bibfield  {author} {\bibinfo {author} {\bibfnamefont {K.~N.}\ \bibnamefont
  {Abazajian}},\ }\href@noop {} {\bibfield  {journal} {\bibinfo  {journal}
  {Phys. Rev. Lett.}\ }\textbf {\bibinfo {volume} {112}},\ \bibinfo {pages}
  {161303} (\bibinfo {year} {2014})}\BibitemShut {NoStop}%
\bibitem [{\citenamefont {Bozek}\ \emph {et~al.}(2016)\citenamefont {Bozek},
  \citenamefont {Boylan-Kolchin}, \citenamefont {Horiuchi}, \citenamefont
  {Garrison-Kimmel}, \citenamefont {Abazajian},\ and\ \citenamefont
  {Bullock}}]{bozek2016}%
  \BibitemOpen
  \bibfield  {author} {\bibinfo {author} {\bibfnamefont {B.}~\bibnamefont
  {Bozek}}, \bibinfo {author} {\bibfnamefont {M.}~\bibnamefont
  {Boylan-Kolchin}}, \bibinfo {author} {\bibfnamefont {S.}~\bibnamefont
  {Horiuchi}}, \bibinfo {author} {\bibfnamefont {S.}~\bibnamefont
  {Garrison-Kimmel}}, \bibinfo {author} {\bibfnamefont {K.}~\bibnamefont
  {Abazajian}}, \ and\ \bibinfo {author} {\bibfnamefont {J.~S.}\ \bibnamefont
  {Bullock}},\ }\href@noop {} {\bibfield  {journal} {\bibinfo  {journal} {Mon.
  Not. R. Astron. Soc.}\ }\textbf {\bibinfo {volume} {459}},\ \bibinfo {pages}
  {1489} (\bibinfo {year} {2016})}\BibitemShut {NoStop}%
\bibitem [{\citenamefont {Horiuchi}\ \emph {et~al.}(2016)\citenamefont
  {Horiuchi}, \citenamefont {Bozek}, \citenamefont {Abazajian}, \citenamefont
  {Boylan-Kolchin}, \citenamefont {Bullock}, \citenamefont {Garrison-Kimmel},\
  and\ \citenamefont {Onorbe}}]{horiuchi2016}%
  \BibitemOpen
  \bibfield  {author} {\bibinfo {author} {\bibfnamefont {S.}~\bibnamefont
  {Horiuchi}}, \bibinfo {author} {\bibfnamefont {B.}~\bibnamefont {Bozek}},
  \bibinfo {author} {\bibfnamefont {K.~N.}\ \bibnamefont {Abazajian}}, \bibinfo
  {author} {\bibfnamefont {M.}~\bibnamefont {Boylan-Kolchin}}, \bibinfo
  {author} {\bibfnamefont {J.~S.}\ \bibnamefont {Bullock}}, \bibinfo {author}
  {\bibfnamefont {S.}~\bibnamefont {Garrison-Kimmel}}, \ and\ \bibinfo {author}
  {\bibfnamefont {J.}~\bibnamefont {Onorbe}},\ }\href@noop {} {\bibfield
  {journal} {\bibinfo  {journal} {Mon. Not. R. Astron. Soc.}\ }\textbf
  {\bibinfo {volume} {456}},\ \bibinfo {pages} {4346} (\bibinfo {year}
  {2016})}\BibitemShut {NoStop}%
\bibitem [{\citenamefont {Johns}\ \emph {et~al.}(2016)\citenamefont {Johns},
  \citenamefont {Mina}, \citenamefont {Cirigliano}, \citenamefont {Paris},\
  and\ \citenamefont {Fuller}}]{johns2016}%
  \BibitemOpen
  \bibfield  {author} {\bibinfo {author} {\bibfnamefont {L.}~\bibnamefont
  {Johns}}, \bibinfo {author} {\bibfnamefont {M.}~\bibnamefont {Mina}},
  \bibinfo {author} {\bibfnamefont {V.}~\bibnamefont {Cirigliano}}, \bibinfo
  {author} {\bibfnamefont {M.~W.}\ \bibnamefont {Paris}}, \ and\ \bibinfo
  {author} {\bibfnamefont {G.~M.}\ \bibnamefont {Fuller}},\ }\href {\doibase
  10.1103/PhysRevD.94.083505} {\bibfield  {journal} {\bibinfo  {journal} {Phys.
  Rev. D}\ }\textbf {\bibinfo {volume} {94}},\ \bibinfo {pages} {083505}
  (\bibinfo {year} {2016})}\BibitemShut {NoStop}%
\bibitem [{\citenamefont {{Duan}}\ \emph {et~al.}(2008)\citenamefont {{Duan}},
  \citenamefont {{Fuller}}, \citenamefont {{Carlson}},\ and\ \citenamefont
  {{Qian}}}]{duan2008b}%
  \BibitemOpen
  \bibfield  {author} {\bibinfo {author} {\bibfnamefont {H.}~\bibnamefont
  {{Duan}}}, \bibinfo {author} {\bibfnamefont {G.~M.}\ \bibnamefont
  {{Fuller}}}, \bibinfo {author} {\bibfnamefont {J.}~\bibnamefont {{Carlson}}},
  \ and\ \bibinfo {author} {\bibfnamefont {Y.-Z.}\ \bibnamefont {{Qian}}},\
  }\href {\doibase 10.1103/PhysRevLett.100.021101} {\bibfield  {journal}
  {\bibinfo  {journal} {Phys. Rev. Lett.}\ }\textbf {\bibinfo {volume} {100}},\
  \bibinfo {pages} {021101} (\bibinfo {year} {2008})}\BibitemShut {NoStop}%
\bibitem [{\citenamefont {{Dasgupta}}\ \emph {et~al.}(2008)\citenamefont
  {{Dasgupta}}, \citenamefont {{Dighe}}, \citenamefont {{Mirizzi}},\ and\
  \citenamefont {{Raffelt}}}]{dasgupta2008c}%
  \BibitemOpen
  \bibfield  {author} {\bibinfo {author} {\bibfnamefont {B.}~\bibnamefont
  {{Dasgupta}}}, \bibinfo {author} {\bibfnamefont {A.}~\bibnamefont {{Dighe}}},
  \bibinfo {author} {\bibfnamefont {A.}~\bibnamefont {{Mirizzi}}}, \ and\
  \bibinfo {author} {\bibfnamefont {G.~G.}\ \bibnamefont {{Raffelt}}},\ }\href
  {\doibase 10.1103/PhysRevD.77.113007} {\bibfield  {journal} {\bibinfo
  {journal} {Phys. Rev. D}\ }\textbf {\bibinfo {volume} {77}},\ \bibinfo
  {pages} {113007} (\bibinfo {year} {2008})}\BibitemShut {NoStop}%
\bibitem [{\citenamefont {Lunardini}\ \emph {et~al.}(2008)\citenamefont
  {Lunardini}, \citenamefont {M\"uller},\ and\ \citenamefont
  {Janka}}]{lunardini2008}%
  \BibitemOpen
  \bibfield  {author} {\bibinfo {author} {\bibfnamefont {C.}~\bibnamefont
  {Lunardini}}, \bibinfo {author} {\bibfnamefont {B.}~\bibnamefont {M\"uller}},
  \ and\ \bibinfo {author} {\bibfnamefont {H.-T.}\ \bibnamefont {Janka}},\
  }\href {\doibase 10.1103/PhysRevD.78.023016} {\bibfield  {journal} {\bibinfo
  {journal} {Phys. Rev. D}\ }\textbf {\bibinfo {volume} {78}},\ \bibinfo
  {pages} {023016} (\bibinfo {year} {2008})}\BibitemShut {NoStop}%
\bibitem [{\citenamefont {Cherry}\ \emph {et~al.}(2010)\citenamefont {Cherry},
  \citenamefont {Fuller}, \citenamefont {Carlson}, \citenamefont {Duan},\ and\
  \citenamefont {Qian}}]{cherry2010}%
  \BibitemOpen
  \bibfield  {author} {\bibinfo {author} {\bibfnamefont {J.~F.}\ \bibnamefont
  {Cherry}}, \bibinfo {author} {\bibfnamefont {G.~M.}\ \bibnamefont {Fuller}},
  \bibinfo {author} {\bibfnamefont {J.}~\bibnamefont {Carlson}}, \bibinfo
  {author} {\bibfnamefont {H.}~\bibnamefont {Duan}}, \ and\ \bibinfo {author}
  {\bibfnamefont {Y.-Z.}\ \bibnamefont {Qian}},\ }\href {\doibase
  10.1103/PhysRevD.82.085025} {\bibfield  {journal} {\bibinfo  {journal} {Phys.
  Rev. D}\ }\textbf {\bibinfo {volume} {82}},\ \bibinfo {pages} {085025}
  (\bibinfo {year} {2010})}\BibitemShut {NoStop}%
\bibitem [{\citenamefont {Duan}\ \emph {et~al.}(2006)\citenamefont {Duan},
  \citenamefont {Fuller}, \citenamefont {Carlson},\ and\ \citenamefont
  {Qian}}]{duan2006}%
  \BibitemOpen
  \bibfield  {author} {\bibinfo {author} {\bibfnamefont {H.}~\bibnamefont
  {Duan}}, \bibinfo {author} {\bibfnamefont {G.~M.}\ \bibnamefont {Fuller}},
  \bibinfo {author} {\bibfnamefont {J.}~\bibnamefont {Carlson}}, \ and\
  \bibinfo {author} {\bibfnamefont {Y.-Z.}\ \bibnamefont {Qian}},\ }\href@noop
  {} {\bibfield  {journal} {\bibinfo  {journal} {Phys. Rev. D}\ }\textbf
  {\bibinfo {volume} {74}},\ \bibinfo {pages} {105014} (\bibinfo {year}
  {2006})}\BibitemShut {NoStop}%
\bibitem [{\citenamefont {Mirizzi}\ and\ \citenamefont
  {Tom\`as}(2011)}]{mirizzi2011}%
  \BibitemOpen
  \bibfield  {author} {\bibinfo {author} {\bibfnamefont {A.}~\bibnamefont
  {Mirizzi}}\ and\ \bibinfo {author} {\bibfnamefont {R.}~\bibnamefont
  {Tom\`as}},\ }\href {\doibase 10.1103/PhysRevD.84.033013} {\bibfield
  {journal} {\bibinfo  {journal} {Phys. Rev. D}\ }\textbf {\bibinfo {volume}
  {84}},\ \bibinfo {pages} {033013} (\bibinfo {year} {2011})}\BibitemShut
  {NoStop}%
\bibitem [{\citenamefont {N{\"o}tzold}\ and\ \citenamefont
  {Raffelt}(1988)}]{notzold1988}%
  \BibitemOpen
  \bibfield  {author} {\bibinfo {author} {\bibfnamefont {D.}~\bibnamefont
  {N{\"o}tzold}}\ and\ \bibinfo {author} {\bibfnamefont {G.}~\bibnamefont
  {Raffelt}},\ }\href@noop {} {\bibfield  {journal} {\bibinfo  {journal} {Nucl.
  Phys.}\ }\textbf {\bibinfo {volume} {B307}},\ \bibinfo {pages} {924}
  (\bibinfo {year} {1988})}\BibitemShut {NoStop}%
\bibitem [{\citenamefont {{Duan}}\ \emph {et~al.}(2006)\citenamefont {{Duan}},
  \citenamefont {{Fuller}},\ and\ \citenamefont {{Qian}}}]{duan2006c}%
  \BibitemOpen
  \bibfield  {author} {\bibinfo {author} {\bibfnamefont {H.}~\bibnamefont
  {{Duan}}}, \bibinfo {author} {\bibfnamefont {G.~M.}\ \bibnamefont
  {{Fuller}}}, \ and\ \bibinfo {author} {\bibfnamefont {Y.-Z.}\ \bibnamefont
  {{Qian}}},\ }\href {\doibase 10.1103/PhysRevD.74.123004} {\bibfield
  {journal} {\bibinfo  {journal} {Phys. Rev. D}\ }\textbf {\bibinfo {volume}
  {74}},\ \bibinfo {pages} {123004} (\bibinfo {year} {2006})}\BibitemShut
  {NoStop}%
\bibitem [{\citenamefont {Sigl}\ and\ \citenamefont
  {Raffelt}(1993)}]{sigl1993}%
  \BibitemOpen
  \bibfield  {author} {\bibinfo {author} {\bibfnamefont {G.}~\bibnamefont
  {Sigl}}\ and\ \bibinfo {author} {\bibfnamefont {G.}~\bibnamefont {Raffelt}},\
  }\href@noop {} {\bibfield  {journal} {\bibinfo  {journal} {Nucl. Phys.}\
  }\textbf {\bibinfo {volume} {B406}},\ \bibinfo {pages} {423} (\bibinfo {year}
  {1993})}\BibitemShut {NoStop}%
\bibitem [{\citenamefont {Strack}\ and\ \citenamefont
  {Burrows}(2005)}]{strack2005}%
  \BibitemOpen
  \bibfield  {author} {\bibinfo {author} {\bibfnamefont {P.}~\bibnamefont
  {Strack}}\ and\ \bibinfo {author} {\bibfnamefont {A.}~\bibnamefont
  {Burrows}},\ }\href {\doibase 10.1103/PhysRevD.71.093004} {\bibfield
  {journal} {\bibinfo  {journal} {Phys. Rev. D}\ }\textbf {\bibinfo {volume}
  {71}},\ \bibinfo {pages} {093004} (\bibinfo {year} {2005})}\BibitemShut
  {NoStop}%
\bibitem [{\citenamefont {Vlasenko}\ \emph {et~al.}(2014)\citenamefont
  {Vlasenko}, \citenamefont {Fuller},\ and\ \citenamefont
  {Cirigliano}}]{vlasenko2014}%
  \BibitemOpen
  \bibfield  {author} {\bibinfo {author} {\bibfnamefont {A.}~\bibnamefont
  {Vlasenko}}, \bibinfo {author} {\bibfnamefont {G.~M.}\ \bibnamefont
  {Fuller}}, \ and\ \bibinfo {author} {\bibfnamefont {V.}~\bibnamefont
  {Cirigliano}},\ }\href@noop {} {\bibfield  {journal} {\bibinfo  {journal}
  {Phys. Rev. D}\ }\textbf {\bibinfo {volume} {89}},\ \bibinfo {pages} {105004}
  (\bibinfo {year} {2014})}\BibitemShut {NoStop}%
\bibitem [{\citenamefont {Blaschke}\ and\ \citenamefont
  {Cirigliano}(2016)}]{blaschke2016}%
  \BibitemOpen
  \bibfield  {author} {\bibinfo {author} {\bibfnamefont {D.~N.}\ \bibnamefont
  {Blaschke}}\ and\ \bibinfo {author} {\bibfnamefont {V.}~\bibnamefont
  {Cirigliano}},\ }\href@noop {} {\bibfield  {journal} {\bibinfo  {journal}
  {Phys. Rev. D}\ }\textbf {\bibinfo {volume} {94}},\ \bibinfo {pages} {033009}
  (\bibinfo {year} {2016})}\BibitemShut {NoStop}%
\bibitem [{\citenamefont {Stodolsky}(1987)}]{stodolsky1987}%
  \BibitemOpen
  \bibfield  {author} {\bibinfo {author} {\bibfnamefont {L.}~\bibnamefont
  {Stodolsky}},\ }\href@noop {} {\bibfield  {journal} {\bibinfo  {journal}
  {Phys. Rev. D}\ }\textbf {\bibinfo {volume} {36}},\ \bibinfo {pages} {2273}
  (\bibinfo {year} {1987})}\BibitemShut {NoStop}%
\bibitem [{\citenamefont {Thomson}(1992)}]{thomson1992}%
  \BibitemOpen
  \bibfield  {author} {\bibinfo {author} {\bibfnamefont {M.~J.}\ \bibnamefont
  {Thomson}},\ }\href {\doibase 10.1103/PhysRevA.45.2243} {\bibfield  {journal}
  {\bibinfo  {journal} {Phys. Rev. A}\ }\textbf {\bibinfo {volume} {45}},\
  \bibinfo {pages} {2243} (\bibinfo {year} {1992})}\BibitemShut {NoStop}%
\bibitem [{\citenamefont {Bell}\ \emph {et~al.}(1999)\citenamefont {Bell},
  \citenamefont {Volkas},\ and\ \citenamefont {Wong}}]{bell1999}%
  \BibitemOpen
  \bibfield  {author} {\bibinfo {author} {\bibfnamefont {N.~F.}\ \bibnamefont
  {Bell}}, \bibinfo {author} {\bibfnamefont {R.~R.}\ \bibnamefont {Volkas}}, \
  and\ \bibinfo {author} {\bibfnamefont {Y.~Y.~Y.}\ \bibnamefont {Wong}},\
  }\href@noop {} {\bibfield  {journal} {\bibinfo  {journal} {Phys. Rev. D}\
  }\textbf {\bibinfo {volume} {59}},\ \bibinfo {pages} {113001} (\bibinfo
  {year} {1999})}\BibitemShut {NoStop}%
\end{thebibliography}%

\end{document}